\documentclass[aps,prl,preprint,superscriptaddress,floatfix,showpacs]{revtex4-1}

\usepackage{graphicx}
\usepackage{upgreek}
\usepackage{siunitx}
\usepackage{units}
\usepackage{textcomp}
\usepackage[yyyymmdd,hhmmss]{datetime}
\usepackage{amsmath}
\usepackage{amssymb}
\usepackage{bpchem}
 \usepackage{placeins}
\usepackage{xcolor}
\usepackage[draft = true]{hyperref}
\definecolor{link}{rgb}{0,0,0.9}
\definecolor{cite}{rgb}{0,0,0.9}
\definecolor{url}{rgb}{0,0,0.9}
\hypersetup
{
    colorlinks=true,
    linkcolor=link,
    citecolor=cite,
    urlcolor=url,
}

\usepackage{soul}
\definecolor{disc}{rgb}{0.9,1,0.9}

\definecolor{new}{rgb}{1,1,0.9}

\newcommand{\abs}[1]{\left|#1\right|}

\begin{document}


\title{Fully on-chip single-photon Hanbury-Brown and Twiss experiment on a monolithic semiconductor-superconductor platform}

\author{Mario Schwartz}
\thanks{These authors contributed equally to this work}
\email{m.schwartz@ihfg.uni-stuttgart.de}
\affiliation{Institut f\"ur Halbleiteroptik und Funktionelle Grenzfl\"achen, Center for Integrated Quantum Science and Technology (IQ{$^{ST}$}) and SCoPE, University of Stuttgart, Allmandring 3, 70569 Stuttgart, Germany}
\author{Ekkehart Schmidt}
\thanks{These authors contributed equally to this work}
\email{m.schwartz@ihfg.uni-stuttgart.de}
\affiliation{Institute of Micro- and Nanoelectronic Systems, Karlsruhe Institute of Technology (KIT), Hertzstrasse 16, 76187 Karlsruhe, Germany}
\author{Ulrich Rengstl}\affiliation{Institut f\"ur Halbleiteroptik und Funktionelle Grenzfl\"achen, Center for Integrated Quantum Science and Technology (IQ{$^{ST}$}) and SCoPE, University of Stuttgart, Allmandring 3, 70569 Stuttgart, Germany}
\author{Florian Hornung}\affiliation{Institut f\"ur Halbleiteroptik und Funktionelle Grenzfl\"achen, Center for Integrated Quantum Science and Technology (IQ{$^{ST}$}) and SCoPE, University of Stuttgart, Allmandring 3, 70569 Stuttgart, Germany}
\author{Stefan Hepp}\affiliation{Institut f\"ur Halbleiteroptik und Funktionelle Grenzfl\"achen, Center for Integrated Quantum Science and Technology (IQ{$^{ST}$}) and SCoPE, University of Stuttgart, Allmandring 3, 70569 Stuttgart, Germany}
\author{Simone L. Portalupi}\affiliation{Institut f\"ur Halbleiteroptik und Funktionelle Grenzfl\"achen, Center for Integrated Quantum Science and Technology (IQ{$^{ST}$}) and SCoPE, University of Stuttgart, Allmandring 3, 70569 Stuttgart, Germany}
\author{Konstantin Ilin}
\affiliation{Institute of Micro- and Nanoelectronic Systems, Karlsruhe Institute of Technology (KIT), Hertzstrasse 16, 76187 Karlsruhe, Germany}
\author{Michael Jetter}\affiliation{Institut f\"ur Halbleiteroptik und Funktionelle Grenzfl\"achen, Center for Integrated Quantum Science and Technology (IQ{$^{ST}$}) and SCoPE, University of Stuttgart, Allmandring 3, 70569 Stuttgart, Germany}
\author{Michael Siegel}
\homepage[]{www.ims.kit.edu}
\affiliation{Institute of Micro- and Nanoelectronic Systems, Karlsruhe Institute of Technology (KIT), Hertzstrasse 16, 76187 Karlsruhe, Germany} 
\author{Peter Michler}
\email{p.michler@ihfg.uni-stuttgart.de}
\homepage[]{www.ihfg.physik.uni-stuttgart.de}
\affiliation{Institut f\"ur Halbleiteroptik und Funktionelle Grenzfl\"achen, Center for Integrated Quantum Science and Technology (IQ{$^{ST}$}) and SCoPE, University of Stuttgart, Allmandring 3, 70569 Stuttgart, Germany}

\date{\today}

\begin{abstract}
\textbf{Photonic quantum technologies such as quantum cryptography \cite{Gisin.Ribordy.ea.2002}, photonic quantum metrology \cite{7, 8, 9}, photonic quantum simulators and computers \cite{10, 11, Aspuru.Walther.2012} will largely benefit from highly scalable and small footprint quantum photonic circuits. To perform fully on-chip quantum photonic operations, three basic building blocks are required: single-photon sources, photonic circuits and single-photon detectors \cite{Dietrich.Fiore.ea.2016}. Highly integrated quantum photonic chips on silicon and related platforms have been demonstrated incorporating only one \cite{3} or two \cite{18} of these basic building blocks. Previous implementations of all three components were mainly limited by laser stray light, making temporal filtering necessary \cite{29} or required complex manipulation to transfer all components onto one chip \cite{32}. So far, a monolithic, simultaneous implementation of all elements demonstrating single-photon operation remains elusive.  Here, we present a fully-integrated Hanbury-Brown and Twiss setup on a micron-sized footprint, consisting of a \BPChem{GaAs} waveguide embedding quantum dots as single-photon sources, a waveguide beamsplitter and two superconducting nanowire single-photon detectors. This enables a second-order correlation measurement at the single-photon level under both continuous-wave and pulsed resonant excitation.
}
\end{abstract}


\maketitle

Up to now, most quantum waveguide (WG) circuits have been fabricated from glass-based and Si-based materials. Both material platforms do not allow monolithic integration of deterministic single-photon sources. The used \BPChem{InGaAs}/\BPChem{GaAs} material system benefits from the capability of directly integrating on-demand non-classical light sources, namely semiconductor quantum dots (QDs) \cite{19}. These emitters reach state-of-the-art performances in terms of single and indistinguishable photon emission, typically via a resonant excitation scheme \cite{Michler.2017}.
Within this platform, single-photon emission in combination with single-mode WGs and beamsplitters (BSs) was demonstrated with and without resonant excitation \cite{21, Sapienza.Thyrrestrup.ea.2010, 23, 30, 31, Schauber.Schall.ea.2018}. Moreover, the implementation of superconducting nanowire single-photon detectors (SNSPDs) was successfully demonstrated on this material system \cite{WGSNSPD, 12, 29}. These detectors represent the most suitable choice for working at the single photon level due to their potential near-unity detection efficiency  (\SI{93}{\percent} \cite{marsili2013detecting}), low dark count rate and very high time resolution with intrinsic timing jitters in the ps range \cite{Jitter1, Jitter2}. 

On the other hand, for silicon and silicon-related quantum photonic platforms a high degree of device complexity was reached, but efficient on-demand non-classical light sources are still missing \cite{18}. By using parametric down conversion sources, only probabilistic single-photon emission is possible and the amount of stray light coming from the intense pump laser prevented so far the implementation of single-photon detectors on the same chip. Electrically-driven sources may solve this issue \cite{32, Benthal.Hallett.ea.2016}, but the used non-resonant excitation scheme typically leads to the emission of photons with a limited degree of indistinguishably that are less suitable for Hong-Ou-Mandel interference based operations. Another method attempted to overcome this limitation relies on the integration of optically active materials on silicon, but this hybrid approach significantly increases the fabrication complexity in comparison to a fully monolithic design \cite{detQDonSi, hetPICQD, Murray.Ellis.ea.2015, Ellis.Bennett.ea.2018}. 

Here, a fully-\BPChem{GaAs}-integrated device is presented (Fig.\,\ref{fig:fig1}(a)). For its characterization a standard cryogenic setup at \SI{4}{\kelvin} with free-space access through optical windows is used. The QDs are embedded into a single-mode \BPChem{GaAs} ridge WG, where individual QDs can be optically excited with a resonant pump laser. The single-photon stream propagates across a 50:50 BS equipped with two independent SNSPDs at the output ports. In the present sample, the laser stray light is suppressed by means of metallic layers, deterministically placed to cover parts of the photonic chip. 
The utilized sample design allows for the comparison of the single-photon source performances both off- and on-chip. The QD emission propagating in the direction of the cleaved sample facet is collected with a microscope objective (Fig.\,\ref{fig:fig2}(a)) and sent to an off-chip spectrometer or an off-chip Hanbury-Brown and Twiss setup.

The emission spectrum of the QD under pulsed resonant excitation (Fig.\,\ref{fig:fig2}(b)) shows very low laser background propagating inside the WG: a ratio of 40:1 for QD emission-to-laser-background is observed. Under continuous wave (cw) excitation, the QD emission-to-stray-light ratio increases to 80:1, due to the smaller linewidth of the excitation laser compared to the linewidth used for pulsed excitation \cite{31}. 
The investigated QD transition stems from an exciton (X) state with a fine structure splitting (FSS) of \SI{4.60}{\giga\hertz} and linewidths of \SI{1.77}{\giga\hertz} and \SI{1.74}{\giga\hertz}, respectively (Fig.\,\ref{fig:fig2}(c)). This is comparable to FSSs and linewidths of QDs in bulk material \cite{richter2010low}, i.e. no degradation of the optical quality of the QD emission due to fabrication processes can be observed. 

The off-chip Hanbury-Brown and Twiss experiment under pulsed resonant excitation shows nearly no coincidences at zero time delay, resulting in a $g^{(2)}(0)$ of \num{0.08(3)} (Fig.\,\ref{fig:fig2}(d)). The small non-vanishing part is caused by the low laser background inside the system and a possible re-excitation of the QD by the \SI{35}{\pico\second} long excitation laser pulses \cite{fischer2016dynamical} in almost equal proportion. In order to directly compare the device performances under off-chip and on-chip detection, no spectral filtering was used in the free-space experiment, since no light filtering is currently implemented on-chip.

On-chip, the emission of the resonantly excited QD  is guided to the coupling region of the on-chip BS, where it gets separated in the two WG arms and guided to the two SNSPDs. The measurements were done using the previously characterized QD. We reach detection efficiencies of up to \SI{47.5(157)}{\percent} for the first (SNSPD-1) and \SI{11.2(38)}{\percent} for the second SNSPD (SNSPD-2) close to their respective critical currents. During the measurements the bias current was set to approximately \SI{90}{\percent} of the critical current to reduce the dark count rate. The associated detection efficiencies of the SNSPDs for these bias levels are \SI{21.8(72)}{\percent} and \SI{1.8(6)}{\percent}.

Time-correlated single-photon counting (TCSPC) experiments under resonant excitation were performed using both SNSPDs revealing a mono-exponential decay with superimposed distinct oscillations (Fig.\,\ref{fig:fig3}(a)).


The decay time measured with SNSPD-1 (SNSPD-2) is \SI{275.5(6)}{\pico\second} (\SI{277.3(21)}{\pico\second}) and the oscillation period \SI{215.2(10)}{\pico\second} (\SI{213.5(8)}{\pico\second}), respectively. This corresponds to \SI{4.647(21)}{\giga\hertz} (\SI{4.683(18)}{\giga\hertz}) FSS of the QD X state. This is found in good agreement with the directly measured spectral detuning of the two fine structure components, which is \SI{4.60(14)}{\giga\hertz} (Fig.\,\ref{fig:fig2}(c)). The depth of the oscillations is only limited by the instrument response function of our detection setup (approximately \SI{100}{\pico\second}). The WG acts as a polarization filter for QD emission with a polarization parallel to the WG since only the perpendicular in-plane polarization mode couples to the WG mode. These so-called quantum beats further prove the coherent nature of the used excitation method \cite{PhotonBeats}. 

Correlating the two signals from the SNSPDs enables fully-integrated on-chip second-order correlation measurements. Under resonant cw excitation a distinct photon anti-bunching dip with a $g^{(2)}(0)$ value of \num{0.24(6)} is found from the raw measurement data (Fig.\,\ref{fig:fig3}(b)). This proves the functionality of a fully-integrated on-chip Hanbury-Brown and Twiss setup including all components in the same device. The laser stray light is strongly reduced, in comparison to previous state-of-the-art experiments, by covering the detector and the WG area with an \BPChem{AlN}/\BPChem{Al} bilayer. The deterministic alignment of the metallic layers shown here turned out to be of key importance for the reliability of the device. By subtracting dark counts, the $g^{(2)}(0)$ value reduces to \num{.17(4)}, which means that over \SI{90}{\percent} of the detected photons are coming from QD emission.
In order to use the QD as an on-demand source of single photons, we switched the excitation scheme to pulsed, still being in resonance with the X state.

A clear suppression of the zero-dealy peak is observed with a $g^{(2)}(0)$ value of \num{0.41(4)}, when the dark counts of the detectors are subtracted (in comparison, a $g^{(2)}(0)$ value of \num{0.59(6)} is found from raw data). With $g^{(2)}(0) < 0.5$ it can be seen that dominant signal stems from the single QD (\SI{77}{\percent}), proving the full operation also under triggered excitation. The higher $g^{(2)}(0)$ value in comparison to the one observed under cw excitation, is caused by the spectrally broader linewidth of the excitation laser and therefore an effectively lower signal-to-noise ratio \cite{31}. Since the intensity of the QD emission drops when switching from cw to pulsed excitation, the relative dark count rate is also higher. 

However, there is a significant scope to increase the signal-to-noise ratio by some obvious modifications of the photonic chip and the operation conditions. We anticipate a reduction of the stray light by further increasing the covered area and putting the Al-cover closer to the WGs. Detector dark counts can be reduced by decreasing the temperature of the used cryogenic setup \cite{kitaygorsky2007dark}. A further reduction of dark counts can be achived by properly shielding the detector from thermal photons. This can conveniently be done by the use of a fiber coupled instead of a free space setup. By placing an identically designed SNSPD in a fiber coupled setup, shielded in a metallic enclosure at \SI{4.2}{\kelvin}, we were able to reduce the number of dark counts by over 2 orders of magnitude (see supplementary). Additionally, the signal level can be increased, e.g. by embedding the QDs into a WG resonator structure, and therefore increasing the coupling efficiency to the guided mode. Furthermore, the WG attenuation could be decreased by lowering the surface roughness through an improved etching process \cite{23} and a removal of the residual AlN-layer on top. 

In conclusion, we successfully demonstrated a fully-integrated on-chip Hanbury-Brown and Twiss experiment, monolithically realized as a  micron-sized \BPChem{GaAs} based device. The single photons emitted by a resonantly excited QD were guided and splitted in a single-mode waveguide BS and then detected on-chip with two \BPChem{NbN} superconducting nanowire single-photon detectors. A effective excitation laser stray light suppression using Al-covers allowed operation without any need of spectral filtering or time gating. The single-photon nature of the emission is clearly proven under cw and pulsed resonant excitation via the on-chip Hanbury-Brown and Twiss setup.
For realizing large-scale quantum photonic chips with advanced functionality, additional elements such as phase shifters and multiple sources of indistinguishable single photons can be straightforward implemented on chip. Both components have already been successfully demonstrated on the InGaAs/GaAs platform. The first one even on-chip \cite{Dietrich.Fiore.ea.2016} and the second one on the basis of wavelength-tunable individual devices \cite{Patel.Bennett.ea.2010}. The wavelength tuning can be enabled e.g. by strain or electric field tuning. The latter could be implemented by applying electrical contacts on both sides of the WGs to use the quantum-confined Stark effect for tuning the emission wavelength of the individual single-photon sources. Such measures together with the presented results pave the way for attractive on-chip quantum photonic applications such as tests of boson sampling, on-chip teleportation and photonic quantum computing.


\FloatBarrier

\section{Methods}
\begin{figure*}[ht]
	\centering
	\includegraphics[width=1\textwidth]{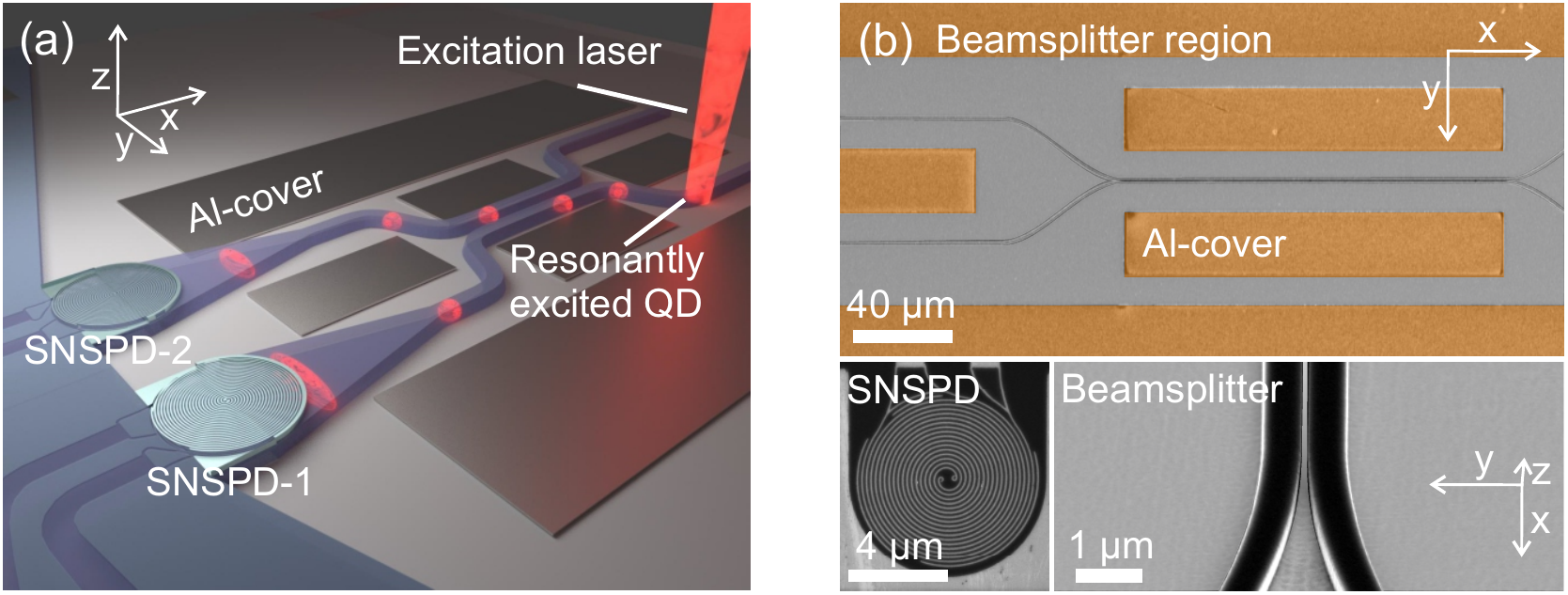}
	\caption{(a): Schematic representation of the device including a resonantly excited QD, a BS, two SNSPDs  and an \BPChem{AlN}/\BPChem{Al} bilayer for covering. The \BPChem{AlN}/\BPChem{Al} bilayer on the SNSPDs is not shown for clarity. (b): False color SEM image of parts of a chip after fabrication of the covers (orange) identical to the one used in this work. Top: Overview picture of the coupler region from top. Left bottom: SEM image of one of the used SNSPDs from top. Right bottom:  Angular view of a zoom in of the coupler region.}
	\label{fig:fig1}
	\end{figure*}

\begin{figure}[ht]
	\centering
	\includegraphics[]{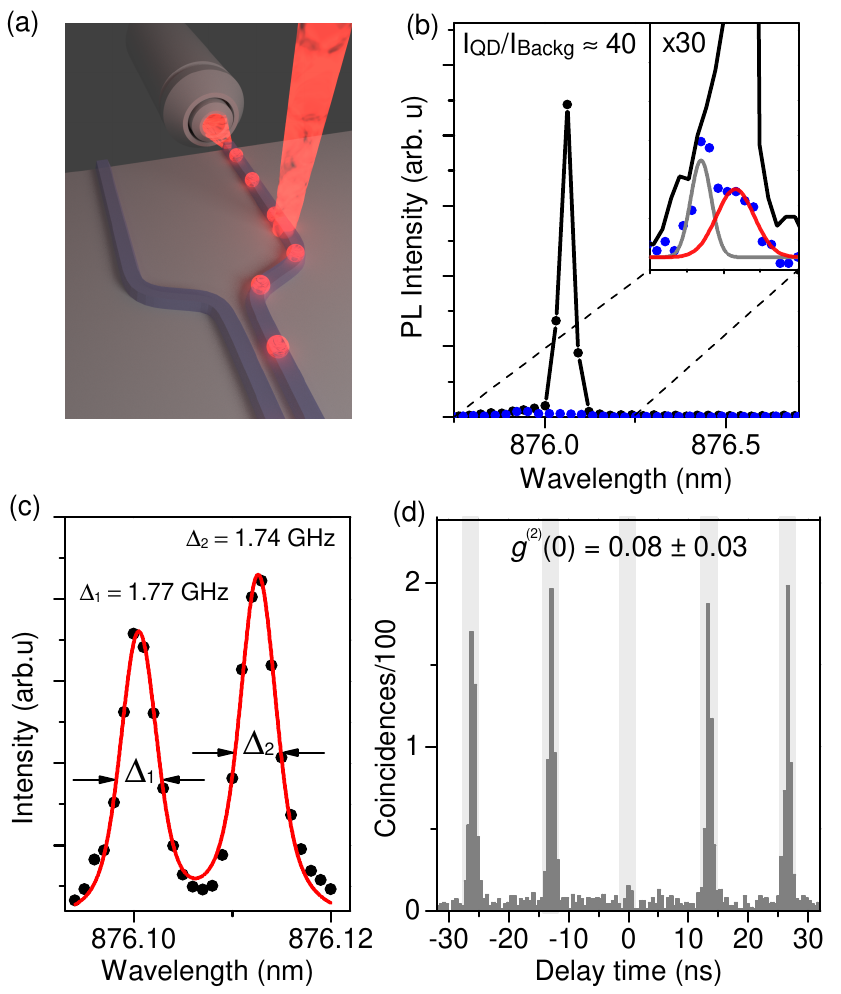}
	\caption{(a) Sketch of the QD excitation and the side collection for off-chip experiments. (b) Emission spectrum of the QD under pulsed resonant excitation. The inset shows a zoom in on the lower part of the emission line. The blue points represent the spectrum when the excitation laser is detuned from the resonant wavelength. The red fit gives the emission-part of the QD caused by the stabilization laser (more information see methods), the gray curve the laser background. (c) Integrated QD intensity under a high-resolution resonant cw laser-scan shows the exciton fine-structure splitting of the investigated state fitted with a double-Voigt profile. (d) Second-order correlation measurement under pulsed resonant excitation. The $g^{(2)}(0)$ value is calculated by dividing the number of coincidences from the peak at zero time delay by the average number of coincidences from the other peaks in time intervals of \SI{2.56}{\nano\second}.}
	\label{fig:fig2}
\end{figure}

\begin{figure}[ht]
	\centering
	\includegraphics[width=1\textwidth]{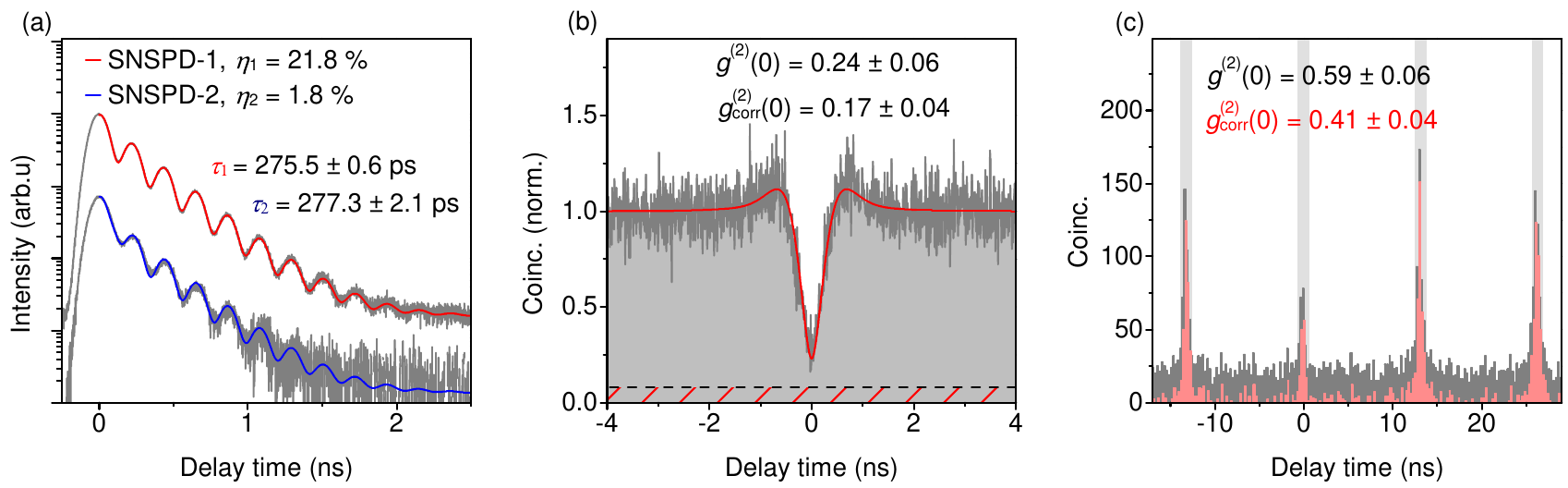}
	\caption{(a) TCSPC measurements of the QD emission measured with SNSPD-1 and SNSPD-2. Both curves exhibit oscillations due to quantum beating between the two fine structure components of the QD state. The fit function is the product of an exponential decay and a harmonic oscillation convolved with the instrumental response. $\eta _1$ and $\eta _2$ depicts the detection efficiencies of the two detectors during measurements. (b) On-chip  correlation measurement via cw resonant excitation. The used fit function includes Rabi oscillations and spectral diffusion as mechanisms for the measured small bunching superimposed to the antibunching at zero time delay. The black dashed line indicates the zero level if dark counts are subtracted. $g^{(2)}_{\text{corr}}(0)$ represents the dark count corrected value. (c) On-chip correlation measurement via pulsed resonant excitation.  
The $g^{(2)}(0)$ value is calculated by dividing the number of coincidences from the peak at zero time delay by the average number of coincidences from the other peaks in time intervals of \SI{1.80}{\nano\second}.}
\label{fig:fig3}
\end{figure}

The chip was grown on a \BPChem{GaAs} (100) substrate using Metal-Organic Vapor-Phase Epitaxy (MOVPE) and is composed of a \SI{2}{\micro\meter} thick \BPChem{Al\_{0.4}Ga\_{0.6}As} confinement layer and a \SI{370}{\nano\meter} thick \BPChem{GaAs} core layer with implemented self-assembled \BPChem{InGaAs}/\BPChem{GaAs} QDs. The design of the sample without on-chip detectors can be found in \cite{23, 31}. The SNSPDs were structured on top out of a \SI{4.2}{\nano\meter} thick \BPChem{NbN} film with a critical temperature of \SI{10.4}{\kelvin}. This value was determined as the temperature were the resistance drops to \SI{0.1}{\percent} of the resistance before its superconducting transition. Between the \BPChem{InGaAs} layer and the \BPChem{NbN}, a \SI{10}{\nano\meter} thick \BPChem{AlN}-buffer layer was introduced to improve \BPChem{NbN}-film properties and therefore the sensitivity of our SNSPDs \cite{schmidt2017aln}. To decrease the effects of current crowding \cite{clem2011geometry}, the SNSPD were shaped as an archimedean spiral in double spiral configuration (Fig.\,\ref{fig:fig1}(a)) \cite{henrich2013detection}. The nanowires have a total length of \SI{218}{\micro\meter} and a width of \SI{115(5)}{\nano\meter} with a gap of \SI{85(5)}{\nano\meter} between the wires: further details on the fabrication of the nitride films and SNSPDs on \BPChem{GaAs} can be found in \cite{schmidt2017aln}. The WG, including the evanescent field coupler, were patterned on the detector chip using electron beam lithography with \SI{50}{\nano\meter} thick hydrogen silsesquioxane (HSQ) resist and an etching process performed via inductively coupled plasma reactive ion etching in Argon + \BPChem{SiCl\_4} atmosphere. After etching, the HSQ and the \BPChem{AlN} buffer layer remained on the WG.
The WGs were patterned to a width of \SI{580(10)}{\nano\meter} and etched to a depth of \SI{320}{\nano\meter} to support only the fundamental TE and TM mode. The 50:50 BS was designed as a directional WG coupler with a length of \SI{150}{\micro\meter} and a gap of \SI{95(5)}{\nano\meter}. Since the detector has a diameter of \SI{8}{\micro\meter}, the WG was broadened using a linear taper over a distance of \SI{250}{\micro\meter}, beginning at a width of approximately \SI{580}{\nano\meter} to a final width of \SI{10}{\micro\meter} to embed the complete detector area (Fig.\,\ref{fig:fig1}(b)). To decrease the effect of direct stray light from the exciting laser, the detectors were covered by \SI{50}{\nano\meter} of \BPChem{AlN} for isolation, followed by a \SI{130(15)}{\nano\meter} thick reflecting \BPChem{Al} layer deposited by sputtering followed by lift-off. The same \BPChem{AlN}/\BPChem{Al} bilayer was added along the sides of the WG to decrease the amount of stray light scattered in the substrate (Fig.\,\ref{fig:fig1}(b)).

After fabrication, the chip was cleaved perpendicularly to the WGs and placed into a liquid helium flow cryostat at a temperature of approximately \SI{4}{\kelvin}. The used optical setup allows free-space access at the top and the side of the sample through quartz windows in the cryostat lid, focusing and collecting the light via microscope objectives. This enables the excitation of QDs from the top with a laser source and the simultaneous observation of light transmitted into both directions of the WG.
In one direction the emission of the QD propagates approximately \SI{400}{\micro\meter} in the straight WG, is coupled out and collected with a microscope objective. In the other direction, the light propagates to an evanescent field coupler, where it is separated into two distinct WG arms. Both arms are tapered at the end and are containing a SNSPD. The overall propagation distance from the QD to the SNSPDs is approximately \SI{1082}{\micro\meter}. The absorption of our on-chip WG was characterized to be \SI{9.29(113)}{\decibel\per\milli\meter}. This value is higher in comparison to state of the art WGs of our group, for which attenuation levels of \SI{2.61(57)}{\decibel\per\milli\meter} were reached \cite{23}. We assign this to a higher surface roughness of the WG sidewalls and absorption in the remaining \BPChem{AlN} buffer layer on top of the WG. Off-chip characterization of cw and pulsed resonant-excitation was performed with one QD in the WG at a wavelength of approximately \SI{876.1}{\nano\meter}. For cw resonant excitation, the spectral width of the excitation laser was $\approx$\,\SI{500}{\mega\hertz}. For pulsed resonant excitation the QD was excited starting with spectrally broad excitation pulses (FWHM $\approx$ \SI{100}{\giga\hertz}). These pulses were shaped in order to achieve a sufficiently small spectral linewidth, closer to the emission linewidth of the QD to further reduce the laser stray light in the system (temporal width $\approx$ \SI{35}{\pico\second}, FWHM $\approx$ \SI{15}{\giga\hertz)} \cite{31}. The polarisation of the excitation laser was chosen parallel to the WG and thereby tilted at around  45° towards the polarisation axes of the X state of the QD. To reduce charge fluctuations in the vicinity of the QD, an additional weak off-resonant laser ($<$ \SI{1}{\percent} power of the resonant laser) at wavelength of \SI{633}{\nano\meter} was used \cite{Nguyen.Sallen.ea.2012}.\\ 
For resonant cw excitation of the QD the overall count rate of SNSPD-1 (SNSPD-2) was \SI{400e3}{\hertz} (\SI{40e3}{\hertz}), the contribution of stray light counts from the excitation and stabilization laser was \SI{23e3}{\hertz}(\SI{6e3}{\hertz}) and the share of dark counts was \SI{250}{\hertz} (\SI{4e3}{\hertz}). Therefore, the ratio between QD emission and noise in form of stray light and dark counts is 16:1 for SNSPD-1 and 3:1 for SNSPD-2. This leads to a theoretical $g^{(2)}(0)$ value of \num{0.29} which is in good agreement to the fitted $g^{(2)}(0)$ value of \num{0.24(6)}. In comparison to measurements without \BPChem{Al} covers (not shown), the amount of straylight decreases by approximately a factor of \num{20}.
\\The data that support the findings of this study are available from the corresponding author upon reasonable request.

\begin{acknowledgments}
\section{Acknowledgments}
The authors would like to thank T. Reindl and E. Reutter in the group of J. Weis from the Max Planck Institute for Solid State Research for electron beam lithography of the waveguide structures. This work was supported by the DFG project MI 500/29-1. The research of the IQST is financially supported by the Ministry of Science, Research and Arts Baden-Württemberg
\end{acknowledgments}
\section{Figures and captions}

\FloatBarrier 
%

\section{Supplementary information to ``Fully on-chip single-photon Hanbury-Brown and Twiss experiment on a monolithic semi-conductor-superconductor platform"}
\renewcommand{\thefigure}{S\arabic{figure}} 
\section{SNSPD characterization}
\setcounter{figure}{0} 

\begin{figure*}[ht]
	\centering
		\includegraphics[width=12cm]{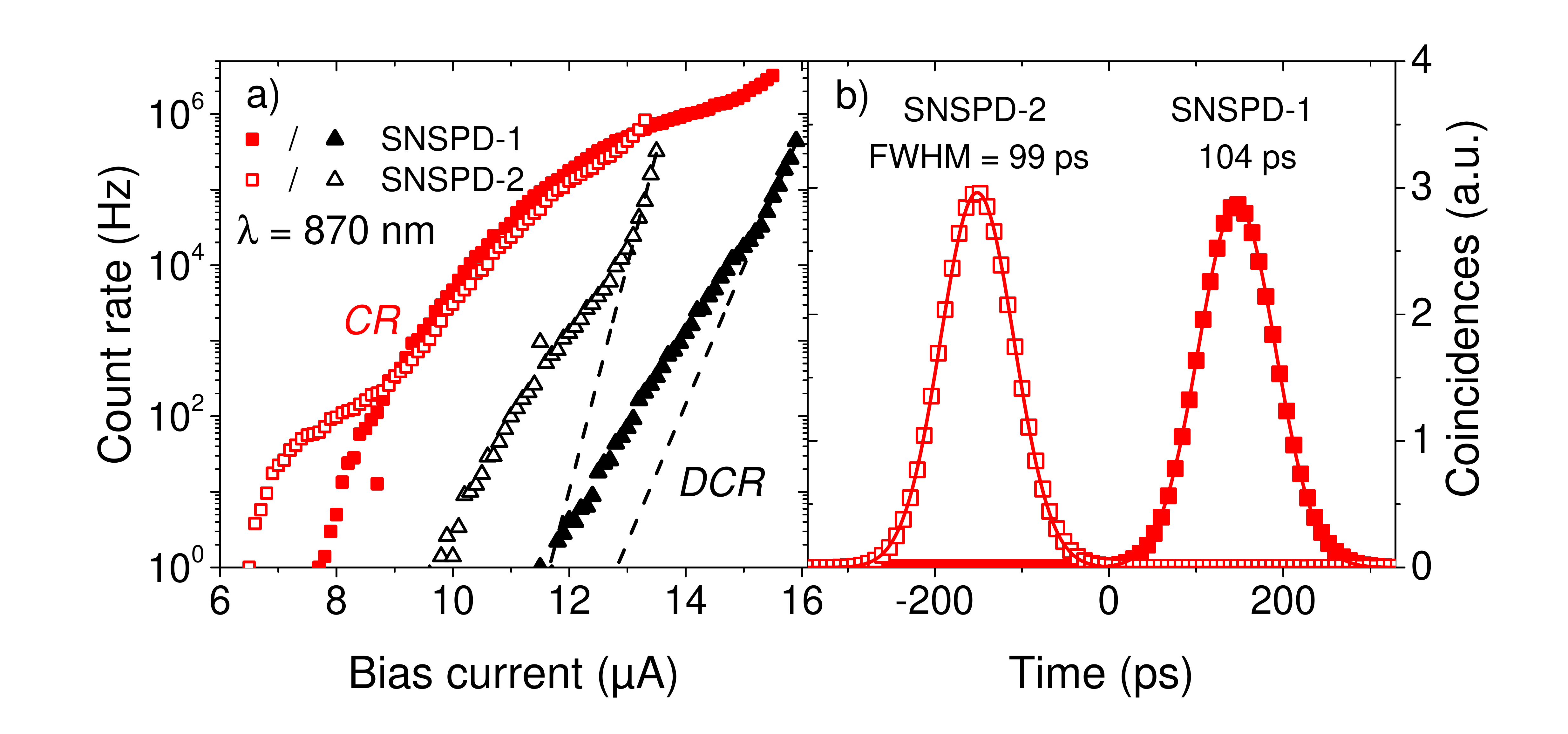}
		\caption{a)      Count rates as a function of the bias current of the detectors. Red squares show the count rates (CRs) when a \SI{870}{\nano\meter} laser was positioned between the two detectors. Black triangles show the dark count rate (DCR) of the SNSPDs in the used free space setup. Results obtained on SNSPD-1 are shown in filled symbols and on SNSPD-2 are in open symbols. Dashed lines show an exponential fit to the exponential trend of the DCR just before the SNSPDs critical current.
b)      Time response of the detected signal for both SNSPDs in correlation to the laser trigger is shown, in filled symbols for SNSPD-1 and in open symbols for SNSPD-2 for a \SI{870}{\nano\meter}  pulsed ps-laser. The corresponding continuous lines are gaussian fits to the measured data.}
	\label{fig:CountRatePlusJitter}
\end{figure*}

The SNSPDs were characterized at \SI{4}{\kelvin} using laser emission at a wavelength of \SI{870}{\nano\meter}, centered between both SNSPDs (Fig. \,\ref{fig:CountRatePlusJitter}(a)). In this way, the SNSPDs can be characterized by utilizing stray light coming from the substrate, since the SNSPDs cannot be illuminated directly from top, due to the AlN/Al cover. SNSPD-1 has a critical current of \SI{15.9}{\micro\ampere} and SNSPD-2 of \SI{13.5}{\micro\ampere}. The critical current of SNSPD-2 is suppressed due to a \SI{50}{\nano\meter} long single constriction where the nanowire width is reduced to \SI{85}{\nano\meter}. The count rate dependencies of both SNSPDs coincide with each other for bias currents between \num{9} and \SI{12}{\micro\ampere}. For lower bias currents the effect of the constriction in the SNSPD-2 is observed. The higher current density of the nanowire at the constriction makes it sensitive for single photons at a bias current of \SI{6.3}{\micro\ampere} and a strong increase in the count rate can be observed. At a bias current of ~\SI{9}{\micro\ampere} the rest of the nanowire is starting to get sensitive for single photons and the full nanowire contributes to the measured count rate \cite{zhang2014characterization}.
To estimate the detection efficiency of the SNSPDs during the experiments, we used the integrated quantum dot under resonantly pulsed excitation. From the power dependent measurements shown in Fig.\,\ref{fig:figS1} we can extract an excited state preparation of \SI{70(1)}{\percent}. In combination with the repetition rate of the laser, the calculated coupling efficiency for QD emission to the WG (\SI{9.8(7)}{\percent}) and the propagation losses over a distance of \SI{1082}{\micro\meter}, a detection efficiency of \SI{21.8(72)}{\percent} for SNSPD-1 and \SI{1.8(6)}{\percent} for SNSPD-2 are calculated at the bias currents of \SI{13.5}{\micro\ampere} and \SI{11.9}{\micro\ampere} respectively.
Taking into account the bias dependence of the detectors (Fig.\,\ref{fig:CountRatePlusJitter}(a)) we can calculate the efficiencies of our detectors to \SI{47.5(157)}{\percent} at a bias current of \SI{14.9}{\micro\ampere} for SNSPD-1, the detection efficiency of SNSPD-2 is limited to \SI{11.2(38)}{\percent} at a bias current of \SI{13.3}{\micro\ampere}, due to the constriction.
The dark count rates (DCR) show clear signs of parasitic counts. When decreasing the bias currents below \SI{15.2}{\micro\ampere} for SNSPD-1 and \SI{13}{\micro\ampere} for SNSPD-2 respectively, it deviates from an exponential bias dependence, which would be expected for dark counts intrinsic to a SNSPD \cite{yamashita2011origin}. One of the reasons for this parasitic counts can be due to thermal radiation due to the open nature of our used free space cryostat. Furthermore, SNSPD-2 suffers from a high number of dark counts caused by the higher current density in the constriction. A comparison of DCRs between the used free space setup and a setup where an SNSPD with similar design was optically coupled with a fiber and placed in a metallic enclosure at \SI{4.2}{\kelvin}, shows that in the latter case, the DCR can be reduced by over two orders of magnitude (see Fig.\,\ref{fig:Setup1vsSetup2DCR}).

\section{Instrumental response function}

The instrumental response function ($IRF$) in Fig.\,\ref{fig:CountRatePlusJitter}(b) was measured for the full on-chip detection setup. $IRF_{Single}$ (Eq.:\ref{IRF_SNSPD}) is the $IRF$ for the correlation of a single SNSPD with the laser trigger, as used for TCSPC measurements. It is the convolution of the timing characteristics of the individual components in our setup: The timing variations of the corresponding electrical trigger signal $\tau_{trigger}$, the timing jitter of the corresponding detector $\tau_{SNSPD}$, caused by a the timing inaccuracy between absorption of the photon and occurrence of the detectors voltage pulse, the timing characteristic of the amplifiers and electronic cables $\tau_{amp}$, and the timing resolution of the used TCSPC electronic $\tau_{HydraHarp}$. A Gaussian fit reveals a full width half maximum (FWHM) value of \SI{104}{\pico\second} for SNSPD-1 and \SI{99}{\pico\second} for SNSPD-2 at a bias current of \SI{12}{\micro\ampere}. Due to their very small pulse duration of approx. \SI{3.3}{\pico\second} the influence of the unshaped excitation laser pulses are not taken into account.

$IRF_{System}$ (Eq.:\ref{IRF_System}) is the $IRF$  for a correlation experiment using one SNSPD as start and the other as stop. It is the convolution of the excitation pulse duration $\tau_{laser}$, the timing accuracy of the corresponding detectors $\tau_{SNSPD-1}$ and $\tau_{SNSPD-2}$, and the timing accuracy of the amplifiers and electronic cables $\tau_{amp}$ as well as the used TCSPC electronic $\tau_{HydraHarp}$ for each individual readout channel. We measured $IRF_{System}$ to be \SI{130}{\pico\second} (FWHM). 
\begin{align} \label{IRF_SNSPD}
	\begin{split}
		IRF_{Single} = (\tau_{trigger}^{2}+\tau_{SNSPD}^{2} 
		+ \tau_{amp}^{2}+\tau_{HydraHarp}^{2})^{1/2} = 101 \pm 3 \; ps
	\end{split}
\end{align}
\begin{align} \label{IRF_System}
	\begin{split}
		IRF_{System} =  (\tau_{SNSPD-1}^{2}+\tau_{SNSPD-2}^{2}+ 
		 2*\tau_{amp}^{2}+2*\tau_{HydraHarp}^{2})^{1/2} = 130 \; ps
	\end{split}
\end{align}

\section{Beam splitter}
The used beamsplitter is an evanescent field coupler with a wavelength-dependent  splitting ratio as depicted in Fig.\,\ref{fig:figS2}. The measurements were done by exciting a QD ensemble in the WG with an excitation laser at \SI{800}{\nano\meter}. The photons from the quantum dot ensemble propagated to the beamsplitter and split into both output arms. The emission from the two output ports was collected with a microscope objective and sent to a spectrometer subsequently. Sharp emission lines representing individual quantum dots in the two recorded spectra were fitted with Gaussian functions and compared by their areas. With this method the splitting ratio of the investigated QD emission line of the main text can be given: $56.7/43.3\pm4.4\%$ (A splitting ratio of 0 would mean that no photons are coupled from the WG arm in which the QD is located to the other WG arm).

\section{Resonant excitation}

\begin{figure*}[h]
	\centering
	\includegraphics[width=10cm]{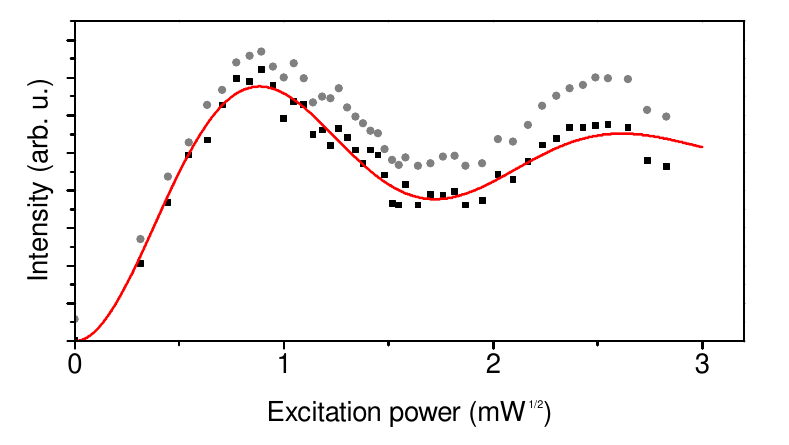}
	\caption{Integrated QD intensity as a function of the square root of the excitation laser power, measured off-chip. The gray circles include the laser stray light, for the black squares it is subtracted.}
	\label{fig:figS1}
\end{figure*}

\begin{figure*}[ht]
	\centering
		\includegraphics[width=10cm]{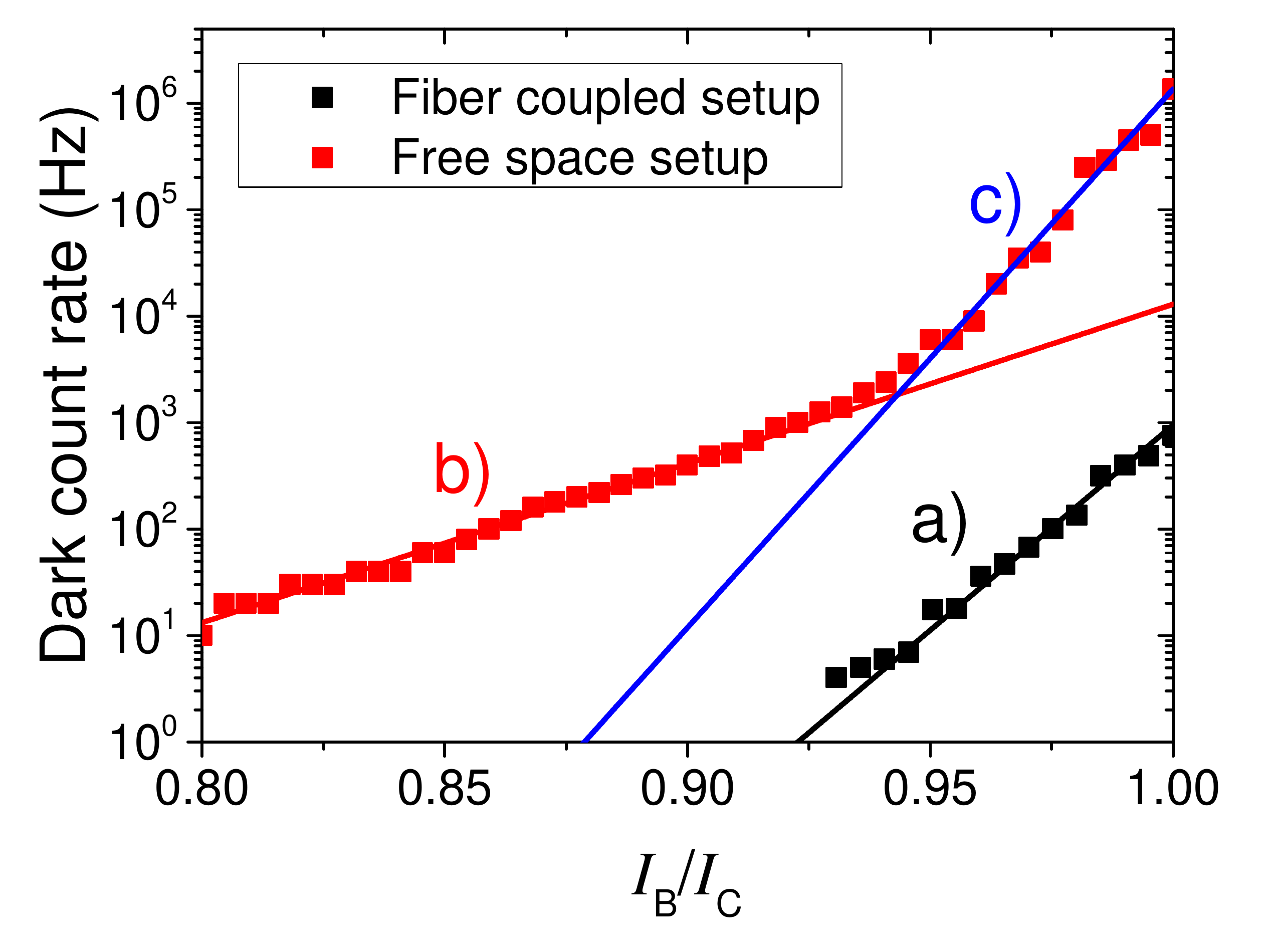}
	\caption{Dark count rate (DCR) on relative bias current $I_B/I_C$ of a similar SNSPD in the used free space setup(red) at \SI{4}{\kelvin} compared to its DCR in a fiber coupled setup(black) at \SI{4.2}{\kelvin}. The lines are exponential fits to different regions of the curve. a) is a fit to the DCR in the fiber coupled setup. b) is a fit to the DCR of the free space setup below \SI{93}{\percent} of the critical current $I_C$  and consists mainly of thermally activated counts. c) is a fit to the DCR of the free space setup above \SI{95}{\percent} of the critical current $I_C$  and consists mainly of intrinsic dark counts.}
	\label{fig:Setup1vsSetup2DCR}
\end{figure*}

\begin{figure*}[ht]
	\centering
	\includegraphics[width=10cm]{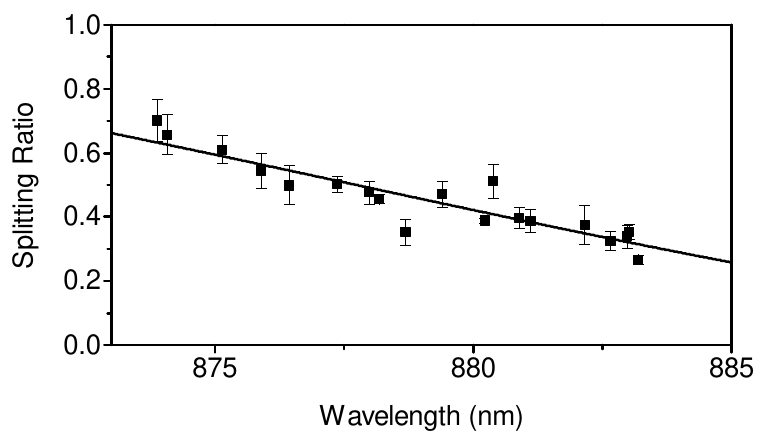}
	\caption{Splitting ratio of our integrated beamsplitter as a function of the emission wavelength of the QDs.}
	\label{fig:figS2}
\end{figure*}

Excitation power dependent measurements of the QD intensity via pulsed resonant excitation revealed clear Rabi-oscillations which show the coherent excitation of the QD (Fig.\,\ref{fig:figS1}). The data with subtracted laser background were fitted numerically by solving the optical Bloch equations of a two level system with a power dependent dephasing rate. A state preparation fidelity of $\num{70(1)}\%$ is extracted from the fit.

For the fitting of the decay curves depiced in Fig.\,3(a) of the main text, the product of an exponential decay and an harmonic oscillation 
\begin{align}\label{eq3s}
	\begin{split}
	Fit(t) = A\exp^{\frac{-t+t_0}{\tau}} \cdot \bigl( 1- \cos(2b \pi t + \phi ) \bigr),
	\end{split}
\end{align}
is used \cite{Flissikowski.Hundt.ea.2001}. $A$, $t_0$, $b$ and $\phi$ are fit-parameters. The function is convolved with the instrumental response function, the pulse duration of the shaped excitation laser pulses (\SI{35}{\pico\second}) and combined with an additional Gaussian function to account for the effect of laser stray light in the system.   

The second order correlation function in Fig.\,3\,(b) of the main text is fitted with the function 
\begin{align}\label{eq1s}
	\begin{split}
	g^{(2)}(\tau) = \Bigl(1-A \cdot \exp^{-\frac{\Gamma _1 + \Gamma _2}{2} \cdot \abs{\tau}} \cdot \bigl(\cos(C \tau)) + \\ 
	+ \frac{\Gamma _1 + \Gamma _2}{2C}\sin(C \abs{\tau})\bigr)\Bigr) \cdot \Bigl(1+B*\exp^{(-|\tau|/\tau_b)}\Bigr).
	\end{split}
	\end{align}
	The fit-function includes the effects of spectral diffusion caused by electrical fluctuation in the vicinity of the QD \cite{Rengstl.Schwartz.ea.2015} and coherent driven state oscillations \cite{Makhonin.Dixon.ea.2014}. The fit parameters $\tau _b$ and $B$  are the time-scale and the amount of the bunching which is caused by spectral diffusion. $A$ determines the amount of antibunching. $\Gamma _1$ is the radiative decay rate, $\Gamma _2$ the damping rate, C = $\sqrt{\Omega ^2 - 1/4 \cdot (\Gamma _1 -\Gamma _2)^2}$ and $\Omega$ the Rabi-frequency. $\Gamma _1$ is assumed to be the reciprocal of the measured decay time from the excited state and $\Gamma _2$ can be calculated by excitation power-dependent measurements of the QD intensity, a measurement of the linewidth $\Delta$ and the formulas \cite{Nguyen.Sallen.ea.20122}      
	
\begin{align}\label{eq2s}
	\begin{split}
	I &\propto \frac{\Omega ^2}{\Gamma _1 \Gamma _2 + \Omega ^2} \propto \frac{P}{P + P_0} \\
	\Delta &= 2\hbar\Gamma_2 \sqrt{1 + \frac{\Omega ^2}{\Gamma _1 \Gamma _2}} = 2\hbar\Gamma_2 \sqrt{1 + \frac{P}{P_0}}. 
	\end{split}
\end{align}	
	
$I$ is the intensity of the QD emission, $P$ is the power of the resonant laser and $P_0$ is the saturation-power of the QD emission. With $\Delta = \SI{1.74}{\giga\hertz}$ and $P =0.43 \cdot P_0$ we can calculate $\Gamma _1 = \SI{3.63e9}{\per\second}$, $\Gamma _2 = \SI{4.52e9}{\per\second}$ and $\Omega = \SI{2.66e9}{\per\second}$. The remaining fit parameters are $A$, $B$ and $\tau _b$. By taking also into account the resolution of our measurement set up of \SI{130}{\pico\second} by deconvoluting the data we achieve $g^{(2)}(0) = 0.24 \pm 0.06$ (If the detector resolution is not taken into account $g^{(2)}_{con}(0) = 0.27 \pm 0.06$ would be the result). As it can be seen in Fig.\,3(b) of the main text the calculated function fits very good to the measurement data, including the width of the antibunching dip which is not fitted rather than calculated with $\Gamma _1$ and $\Gamma _2$.


\begin{thebibliography}{39}%
\makeatletter
\providecommand \@ifxundefined [1]{%
 \@ifx{#1\undefined}
}%
\providecommand \@ifnum [1]{%
 \ifnum #1\expandafter \@firstoftwo
 \else \expandafter \@secondoftwo
 \fi
}%
\providecommand \@ifx [1]{%
 \ifx #1\expandafter \@firstoftwo
 \else \expandafter \@secondoftwo
 \fi
}%
\providecommand \natexlab [1]{#1}%
\providecommand \enquote  [1]{``#1''}%
\providecommand \bibnamefont  [1]{#1}%
\providecommand \bibfnamefont [1]{#1}%
\providecommand \citenamefont [1]{#1}%
\providecommand \href@noop [0]{\@secondoftwo}%
\providecommand \href [0]{\begingroup \@sanitize@url \@href}%
\providecommand \@href[1]{\@@startlink{#1}\@@href}%
\providecommand \@@href[1]{\endgroup#1\@@endlink}%
\providecommand \@sanitize@url [0]{\catcode `\\12\catcode `\$12\catcode
  `\&12\catcode `\#12\catcode `\^12\catcode `\_12\catcode `\%12\relax}%
\providecommand \@@startlink[1]{}%
\providecommand \@@endlink[0]{}%
\providecommand \url  [0]{\begingroup\@sanitize@url \@url }%
\providecommand \@url [1]{\endgroup\@href {#1}{\urlprefix }}%
\providecommand \urlprefix  [0]{URL }%
\providecommand \Eprint [0]{\href }%
\providecommand \doibase [0]{http://dx.doi.org/}%
\providecommand \selectlanguage [0]{\@gobble}%
\providecommand \bibinfo  [0]{\@secondoftwo}%
\providecommand \bibfield  [0]{\@secondoftwo}%
\providecommand \translation [1]{[#1]}%
\providecommand \BibitemOpen [0]{}%
\providecommand \bibitemStop [0]{}%
\providecommand \bibitemNoStop [0]{.\EOS\space}%
\providecommand \EOS [0]{\spacefactor3000\relax}%
\providecommand \BibitemShut  [1]{\csname bibitem#1\endcsname}%
\let\auto@bib@innerbib\@empty
\bibitem [{\citenamefont {Gisin}\ \emph {et~al.}(2002)\citenamefont {Gisin},
  \citenamefont {Ribordy}, \citenamefont {Tittel},\ and\ \citenamefont
  {Zbinden}}]{Gisin.Ribordy.ea.2002}%
  \BibitemOpen
  \bibfield  {author} {\bibinfo {author} {\bibfnamefont {N.}~\bibnamefont
  {Gisin}}, \bibinfo {author} {\bibfnamefont {G.}~\bibnamefont {Ribordy}},
  \bibinfo {author} {\bibfnamefont {W.}~\bibnamefont {Tittel}}, \ and\ \bibinfo
  {author} {\bibfnamefont {H.}~\bibnamefont {Zbinden}},\ }\href {\doibase
  10.1103/RevModPhys.74.145} {\bibfield  {journal} {\bibinfo  {journal} {Rev.
  Mod. Phys.}\ }\textbf {\bibinfo {volume} {74}},\ \bibinfo {pages} {145}
  (\bibinfo {year} {2002})}\BibitemShut {NoStop}%
\bibitem [{\citenamefont {Giovannetti}\ \emph {et~al.}(2006)\citenamefont
  {Giovannetti}, \citenamefont {Lloyd},\ and\ \citenamefont {Maccone}}]{7}%
  \BibitemOpen
  \bibfield  {author} {\bibinfo {author} {\bibfnamefont {V.}~\bibnamefont
  {Giovannetti}}, \bibinfo {author} {\bibfnamefont {S.}~\bibnamefont {Lloyd}},
  \ and\ \bibinfo {author} {\bibfnamefont {L.}~\bibnamefont {Maccone}},\ }\href
  {\doibase 10.1103/PhysRevLett.96.010401} {\bibfield  {journal} {\bibinfo
  {journal} {Phys. Rev. Lett.}\ }\textbf {\bibinfo {volume} {96}},\ \bibinfo
  {pages} {010401} (\bibinfo {year} {2006})}\BibitemShut {NoStop}%
\bibitem [{\citenamefont {Dowling}(2008)}]{8}%
  \BibitemOpen
  \bibfield  {author} {\bibinfo {author} {\bibfnamefont {J.~P.}\ \bibnamefont
  {Dowling}},\ }\href {\doibase 10.1080/00107510802091298} {\bibfield
  {journal} {\bibinfo  {journal} {Contemporary Physics}\ }\textbf {\bibinfo
  {volume} {49}},\ \bibinfo {pages} {125} (\bibinfo {year} {2008})}\BibitemShut
  {NoStop}%
\bibitem [{\citenamefont {M\"uller}\ \emph {et~al.}(2017)\citenamefont
  {M\"uller}, \citenamefont {Vural}, \citenamefont {Schneider}, \citenamefont
  {Rastelli}, \citenamefont {Schmidt}, \citenamefont {H\"ofling},\ and\
  \citenamefont {Michler}}]{9}%
  \BibitemOpen
  \bibfield  {author} {\bibinfo {author} {\bibfnamefont {M.}~\bibnamefont
  {M\"uller}}, \bibinfo {author} {\bibfnamefont {H.}~\bibnamefont {Vural}},
  \bibinfo {author} {\bibfnamefont {C.}~\bibnamefont {Schneider}}, \bibinfo
  {author} {\bibfnamefont {A.}~\bibnamefont {Rastelli}}, \bibinfo {author}
  {\bibfnamefont {O.~G.}\ \bibnamefont {Schmidt}}, \bibinfo {author}
  {\bibfnamefont {S.}~\bibnamefont {H\"ofling}}, \ and\ \bibinfo {author}
  {\bibfnamefont {P.}~\bibnamefont {Michler}},\ }\href {\doibase
  10.1103/PhysRevLett.118.257402} {\bibfield  {journal} {\bibinfo  {journal}
  {Phys. Rev. Lett.}\ }\textbf {\bibinfo {volume} {118}},\ \bibinfo {pages}
  {257402} (\bibinfo {year} {2017})}\BibitemShut {NoStop}%
\bibitem [{\citenamefont {Kok}\ \emph {et~al.}(2007)\citenamefont {Kok},
  \citenamefont {Munro}, \citenamefont {Nemoto}, \citenamefont {Ralph},
  \citenamefont {Dowling},\ and\ \citenamefont {Milburn}}]{10}%
  \BibitemOpen
  \bibfield  {author} {\bibinfo {author} {\bibfnamefont {P.}~\bibnamefont
  {Kok}}, \bibinfo {author} {\bibfnamefont {W.~J.}\ \bibnamefont {Munro}},
  \bibinfo {author} {\bibfnamefont {K.}~\bibnamefont {Nemoto}}, \bibinfo
  {author} {\bibfnamefont {T.~C.}\ \bibnamefont {Ralph}}, \bibinfo {author}
  {\bibfnamefont {J.~P.}\ \bibnamefont {Dowling}}, \ and\ \bibinfo {author}
  {\bibfnamefont {G.~J.}\ \bibnamefont {Milburn}},\ }\href {\doibase
  10.1103/RevModPhys.79.135} {\bibfield  {journal} {\bibinfo  {journal} {Rev.
  Mod. Phys.}\ }\textbf {\bibinfo {volume} {79}},\ \bibinfo {pages} {135}
  (\bibinfo {year} {2007})}\BibitemShut {NoStop}%
\bibitem [{\citenamefont {Ladd}\ \emph {et~al.}(2010)\citenamefont {Ladd},
  \citenamefont {Jelezko}, \citenamefont {Laflamme}, \citenamefont {Nakamura},
  \citenamefont {Monroe},\ and\ \citenamefont {O{\textquoteright}Brien}}]{11}%
  \BibitemOpen
  \bibfield  {author} {\bibinfo {author} {\bibfnamefont {T.~D.}\ \bibnamefont
  {Ladd}}, \bibinfo {author} {\bibfnamefont {F.}~\bibnamefont {Jelezko}},
  \bibinfo {author} {\bibfnamefont {R.}~\bibnamefont {Laflamme}}, \bibinfo
  {author} {\bibfnamefont {Y.}~\bibnamefont {Nakamura}}, \bibinfo {author}
  {\bibfnamefont {C.}~\bibnamefont {Monroe}}, \ and\ \bibinfo {author}
  {\bibfnamefont {J.~L.}\ \bibnamefont {O{\textquoteright}Brien}},\ }\href@noop
  {} {\bibfield  {journal} {\bibinfo  {journal} {Nature}\ }\textbf {\bibinfo
  {volume} {464}},\ \bibinfo {pages} {45} (\bibinfo {year} {2010})}\BibitemShut
  {NoStop}%
\bibitem [{\citenamefont {Aspuru-Guzik}\ and\ \citenamefont
  {Walther}(2012)}]{Aspuru.Walther.2012}%
  \BibitemOpen
  \bibfield  {author} {\bibinfo {author} {\bibfnamefont {A.}~\bibnamefont
  {Aspuru-Guzik}}\ and\ \bibinfo {author} {\bibfnamefont {P.}~\bibnamefont
  {Walther}},\ }\href {http://dx.doi.org/10.1038/nphys2253} {\bibfield
  {journal} {\bibinfo  {journal} {Nature Physics}\ }\textbf {\bibinfo {volume}
  {8}},\ \bibinfo {pages} {285} (\bibinfo {year} {2012})}\BibitemShut {NoStop}%
\bibitem [{\citenamefont {Dietrich}\ \emph {et~al.}(2016)\citenamefont
  {Dietrich}, \citenamefont {Fiore}, \citenamefont {Thompson}, \citenamefont
  {Kamp},\ and\ \citenamefont {Höfling}}]{Dietrich.Fiore.ea.2016}%
  \BibitemOpen
  \bibfield  {author} {\bibinfo {author} {\bibfnamefont {C.~P.}\ \bibnamefont
  {Dietrich}}, \bibinfo {author} {\bibfnamefont {A.}~\bibnamefont {Fiore}},
  \bibinfo {author} {\bibfnamefont {M.~G.}\ \bibnamefont {Thompson}}, \bibinfo
  {author} {\bibfnamefont {M.}~\bibnamefont {Kamp}}, \ and\ \bibinfo {author}
  {\bibfnamefont {S.}~\bibnamefont {Höfling}},\ }\href {\doibase
  10.1002/lpor.201500321} {\bibfield  {journal} {\bibinfo  {journal} {Laser \&
  Photonics Reviews}\ }\textbf {\bibinfo {volume} {10}},\ \bibinfo {pages}
  {870} (\bibinfo {year} {2016})}\BibitemShut {NoStop}%
\bibitem [{\citenamefont {Carolan}\ \emph {et~al.}(2015)\citenamefont
  {Carolan}, \citenamefont {Harrold}, \citenamefont {Sparrow}, \citenamefont
  {Mart{\'\i}n-L{\'o}pez}, \citenamefont {Russell}, \citenamefont
  {Silverstone}, \citenamefont {Shadbolt}, \citenamefont {Matsuda},
  \citenamefont {Oguma}, \citenamefont {Itoh}, \citenamefont {Marshall},
  \citenamefont {Thompson}, \citenamefont {Matthews}, \citenamefont
  {Hashimoto}, \citenamefont {O{\textquoteright}Brien},\ and\ \citenamefont
  {Laing}}]{3}%
  \BibitemOpen
  \bibfield  {author} {\bibinfo {author} {\bibfnamefont {J.}~\bibnamefont
  {Carolan}}, \bibinfo {author} {\bibfnamefont {C.}~\bibnamefont {Harrold}},
  \bibinfo {author} {\bibfnamefont {C.}~\bibnamefont {Sparrow}}, \bibinfo
  {author} {\bibfnamefont {E.}~\bibnamefont {Mart{\'\i}n-L{\'o}pez}}, \bibinfo
  {author} {\bibfnamefont {N.~J.}\ \bibnamefont {Russell}}, \bibinfo {author}
  {\bibfnamefont {J.~W.}\ \bibnamefont {Silverstone}}, \bibinfo {author}
  {\bibfnamefont {P.~J.}\ \bibnamefont {Shadbolt}}, \bibinfo {author}
  {\bibfnamefont {N.}~\bibnamefont {Matsuda}}, \bibinfo {author} {\bibfnamefont
  {M.}~\bibnamefont {Oguma}}, \bibinfo {author} {\bibfnamefont
  {M.}~\bibnamefont {Itoh}}, \bibinfo {author} {\bibfnamefont {G.~D.}\
  \bibnamefont {Marshall}}, \bibinfo {author} {\bibfnamefont {M.~G.}\
  \bibnamefont {Thompson}}, \bibinfo {author} {\bibfnamefont {J.~C.~F.}\
  \bibnamefont {Matthews}}, \bibinfo {author} {\bibfnamefont {T.}~\bibnamefont
  {Hashimoto}}, \bibinfo {author} {\bibfnamefont {J.~L.}\ \bibnamefont
  {O{\textquoteright}Brien}}, \ and\ \bibinfo {author} {\bibfnamefont
  {A.}~\bibnamefont {Laing}},\ }\href {\doibase 10.1126/science.aab3642}
  {\bibfield  {journal} {\bibinfo  {journal} {Science}\ }\textbf {\bibinfo
  {volume} {349}},\ \bibinfo {pages} {711} (\bibinfo {year}
  {2015})}\BibitemShut {NoStop}%
\bibitem [{\citenamefont {Wang}\ \emph {et~al.}(2018)\citenamefont {Wang},
  \citenamefont {Paesani}, \citenamefont {Ding}, \citenamefont {Santagati},
  \citenamefont {Skrzypczyk}, \citenamefont {Salavrakos}, \citenamefont {Tura},
  \citenamefont {Augusiak}, \citenamefont {Man{\v c}inska}, \citenamefont
  {Bacco}, \citenamefont {Bonneau}, \citenamefont {Silverstone}, \citenamefont
  {Gong}, \citenamefont {Ac{\'\i}n}, \citenamefont {Rottwitt}, \citenamefont
  {Oxenl{\o}we}, \citenamefont {O{\textquoteright}Brien}, \citenamefont
  {Laing},\ and\ \citenamefont {Thompson}}]{18}%
  \BibitemOpen
  \bibfield  {author} {\bibinfo {author} {\bibfnamefont {J.}~\bibnamefont
  {Wang}}, \bibinfo {author} {\bibfnamefont {S.}~\bibnamefont {Paesani}},
  \bibinfo {author} {\bibfnamefont {Y.}~\bibnamefont {Ding}}, \bibinfo {author}
  {\bibfnamefont {R.}~\bibnamefont {Santagati}}, \bibinfo {author}
  {\bibfnamefont {P.}~\bibnamefont {Skrzypczyk}}, \bibinfo {author}
  {\bibfnamefont {A.}~\bibnamefont {Salavrakos}}, \bibinfo {author}
  {\bibfnamefont {J.}~\bibnamefont {Tura}}, \bibinfo {author} {\bibfnamefont
  {R.}~\bibnamefont {Augusiak}}, \bibinfo {author} {\bibfnamefont
  {L.}~\bibnamefont {Man{\v c}inska}}, \bibinfo {author} {\bibfnamefont
  {D.}~\bibnamefont {Bacco}}, \bibinfo {author} {\bibfnamefont
  {D.}~\bibnamefont {Bonneau}}, \bibinfo {author} {\bibfnamefont {J.~W.}\
  \bibnamefont {Silverstone}}, \bibinfo {author} {\bibfnamefont
  {Q.}~\bibnamefont {Gong}}, \bibinfo {author} {\bibfnamefont {A.}~\bibnamefont
  {Ac{\'\i}n}}, \bibinfo {author} {\bibfnamefont {K.}~\bibnamefont {Rottwitt}},
  \bibinfo {author} {\bibfnamefont {L.~K.}\ \bibnamefont {Oxenl{\o}we}},
  \bibinfo {author} {\bibfnamefont {J.~L.}\ \bibnamefont
  {O{\textquoteright}Brien}}, \bibinfo {author} {\bibfnamefont
  {A.}~\bibnamefont {Laing}}, \ and\ \bibinfo {author} {\bibfnamefont {M.~G.}\
  \bibnamefont {Thompson}},\ }\href {\doibase 10.1126/science.aar7053}
  {\bibfield  {journal} {\bibinfo  {journal} {Science}\ } (\bibinfo {year}
  {2018}),\ 10.1126/science.aar7053}\BibitemShut {NoStop}%
\bibitem [{\citenamefont {Reithmaier}\ \emph {et~al.}(2015)\citenamefont
  {Reithmaier}, \citenamefont {Kaniber}, \citenamefont {Flassig}, \citenamefont
  {Lichtmannecker}, \citenamefont {M\"uller}, \citenamefont {Andrejew},
  \citenamefont {Vučković}, \citenamefont {Gross},\ and\ \citenamefont
  {Finley}}]{29}%
  \BibitemOpen
  \bibfield  {author} {\bibinfo {author} {\bibfnamefont {G.}~\bibnamefont
  {Reithmaier}}, \bibinfo {author} {\bibfnamefont {M.}~\bibnamefont {Kaniber}},
  \bibinfo {author} {\bibfnamefont {F.}~\bibnamefont {Flassig}}, \bibinfo
  {author} {\bibfnamefont {S.}~\bibnamefont {Lichtmannecker}}, \bibinfo
  {author} {\bibfnamefont {K.}~\bibnamefont {M\"uller}}, \bibinfo {author}
  {\bibfnamefont {A.}~\bibnamefont {Andrejew}}, \bibinfo {author}
  {\bibfnamefont {J.}~\bibnamefont {Vučković}}, \bibinfo {author}
  {\bibfnamefont {R.}~\bibnamefont {Gross}}, \ and\ \bibinfo {author}
  {\bibfnamefont {J.~J.}\ \bibnamefont {Finley}},\ }\href {\doibase
  10.1021/acs.nanolett.5b01444} {\bibfield  {journal} {\bibinfo  {journal}
  {Nano Letters}\ }\textbf {\bibinfo {volume} {15}},\ \bibinfo {pages} {5208}
  (\bibinfo {year} {2015})}\BibitemShut {NoStop}%
\bibitem [{\citenamefont {Khasminskaya}\ \emph {et~al.}(2016)\citenamefont
  {Khasminskaya}, \citenamefont {Pyatkov}, \citenamefont {S{\l}owik},
  \citenamefont {Ferrari}, \citenamefont {Kahl}, \citenamefont {Kovalyuk},
  \citenamefont {Rath}, \citenamefont {Vetter}, \citenamefont {Hennrich},
  \citenamefont {Kappes}, \citenamefont {Gol'tsmann}, \citenamefont {Korneev},
  \citenamefont {Rockstuhl}, \citenamefont {Krupke},\ and\ \citenamefont
  {Pernice}}]{32}%
  \BibitemOpen
  \bibfield  {author} {\bibinfo {author} {\bibfnamefont {S.}~\bibnamefont
  {Khasminskaya}}, \bibinfo {author} {\bibfnamefont {F.}~\bibnamefont
  {Pyatkov}}, \bibinfo {author} {\bibfnamefont {K.}~\bibnamefont {S{\l}owik}},
  \bibinfo {author} {\bibfnamefont {S.}~\bibnamefont {Ferrari}}, \bibinfo
  {author} {\bibfnamefont {O.}~\bibnamefont {Kahl}}, \bibinfo {author}
  {\bibfnamefont {V.}~\bibnamefont {Kovalyuk}}, \bibinfo {author}
  {\bibfnamefont {P.}~\bibnamefont {Rath}}, \bibinfo {author} {\bibfnamefont
  {A.}~\bibnamefont {Vetter}}, \bibinfo {author} {\bibfnamefont
  {F.}~\bibnamefont {Hennrich}}, \bibinfo {author} {\bibfnamefont {M.~M.}\
  \bibnamefont {Kappes}}, \bibinfo {author} {\bibfnamefont {G.}~\bibnamefont
  {Gol'tsmann}}, \bibinfo {author} {\bibfnamefont {A.}~\bibnamefont {Korneev}},
  \bibinfo {author} {\bibfnamefont {C.}~\bibnamefont {Rockstuhl}}, \bibinfo
  {author} {\bibfnamefont {R.}~\bibnamefont {Krupke}}, \ and\ \bibinfo {author}
  {\bibfnamefont {W.~H.~P.}\ \bibnamefont {Pernice}},\ }\href
  {http://dx.doi.org/10.1038/nphoton.2016.178} {\bibfield  {journal} {\bibinfo
  {journal} {Nature Photonics}\ }\textbf {\bibinfo {volume} {10}},\ \bibinfo
  {pages} {727} (\bibinfo {year} {2016})}\BibitemShut {NoStop}%
\bibitem [{\citenamefont {Michler}\ \emph {et~al.}(2000)\citenamefont
  {Michler}, \citenamefont {Kiraz}, \citenamefont {Becher}, \citenamefont
  {Schoenfeld}, \citenamefont {Petroff}, \citenamefont {Zhang}, \citenamefont
  {Hu},\ and\ \citenamefont {Imamoglu}}]{19}%
  \BibitemOpen
  \bibfield  {author} {\bibinfo {author} {\bibfnamefont {P.}~\bibnamefont
  {Michler}}, \bibinfo {author} {\bibfnamefont {A.}~\bibnamefont {Kiraz}},
  \bibinfo {author} {\bibfnamefont {C.}~\bibnamefont {Becher}}, \bibinfo
  {author} {\bibfnamefont {W.~V.}\ \bibnamefont {Schoenfeld}}, \bibinfo
  {author} {\bibfnamefont {P.~M.}\ \bibnamefont {Petroff}}, \bibinfo {author}
  {\bibfnamefont {L.}~\bibnamefont {Zhang}}, \bibinfo {author} {\bibfnamefont
  {E.}~\bibnamefont {Hu}}, \ and\ \bibinfo {author} {\bibfnamefont
  {A.}~\bibnamefont {Imamoglu}},\ }\href {\doibase
  10.1126/science.290.5500.2282} {\bibfield  {journal} {\bibinfo  {journal}
  {Science}\ }\textbf {\bibinfo {volume} {290}},\ \bibinfo {pages} {2282}
  (\bibinfo {year} {2000})}\BibitemShut {NoStop}%
\bibitem [{\citenamefont {Michler~(editor)}(2017)}]{Michler.2017}%
  \BibitemOpen
  \bibfield  {author} {\bibinfo {author} {\bibfnamefont {P.}~\bibnamefont
  {Michler~(editor)}},\ }\href {https://www.springer.com/de/book/9783319563770}
  {\emph {\bibinfo {title} {Quantum Dots for Quantum Information
  Technologies}}}\ (\bibinfo  {publisher} {Springer, Berlin},\ \bibinfo {year}
  {2017})\BibitemShut {NoStop}%
\bibitem [{\citenamefont {Prtljaga}\ \emph {et~al.}(2014)\citenamefont
  {Prtljaga}, \citenamefont {Coles}, \citenamefont {O'Hara}, \citenamefont
  {Royall}, \citenamefont {Clarke}, \citenamefont {Fox},\ and\ \citenamefont
  {Skolnick}}]{21}%
  \BibitemOpen
  \bibfield  {author} {\bibinfo {author} {\bibfnamefont {N.}~\bibnamefont
  {Prtljaga}}, \bibinfo {author} {\bibfnamefont {R.}~\bibnamefont {Coles}},
  \bibinfo {author} {\bibfnamefont {J.}~\bibnamefont {O'Hara}}, \bibinfo
  {author} {\bibfnamefont {B.}~\bibnamefont {Royall}}, \bibinfo {author}
  {\bibfnamefont {E.}~\bibnamefont {Clarke}}, \bibinfo {author} {\bibfnamefont
  {A.~M.}\ \bibnamefont {Fox}}, \ and\ \bibinfo {author} {\bibfnamefont
  {M.~S.}\ \bibnamefont {Skolnick}},\ }\href {\doibase 10.1063/1.4883374}
  {\bibfield  {journal} {\bibinfo  {journal} {Applied Physics Letters}\
  }\textbf {\bibinfo {volume} {104}},\ \bibinfo {pages} {231107} (\bibinfo
  {year} {2014})}\BibitemShut {NoStop}%
\bibitem [{\citenamefont {Sapienza}\ \emph {et~al.}(2010)\citenamefont
  {Sapienza}, \citenamefont {Thyrrestrup}, \citenamefont {Stobbe},
  \citenamefont {Garcia}, \citenamefont {Smolka},\ and\ \citenamefont
  {Lodahl}}]{Sapienza.Thyrrestrup.ea.2010}%
  \BibitemOpen
  \bibfield  {author} {\bibinfo {author} {\bibfnamefont {L.}~\bibnamefont
  {Sapienza}}, \bibinfo {author} {\bibfnamefont {H.}~\bibnamefont
  {Thyrrestrup}}, \bibinfo {author} {\bibfnamefont {S.}~\bibnamefont {Stobbe}},
  \bibinfo {author} {\bibfnamefont {P.~D.}\ \bibnamefont {Garcia}}, \bibinfo
  {author} {\bibfnamefont {S.}~\bibnamefont {Smolka}}, \ and\ \bibinfo {author}
  {\bibfnamefont {P.}~\bibnamefont {Lodahl}},\ }\href {\doibase
  10.1126/science.1185080} {\bibfield  {journal} {\bibinfo  {journal}
  {Science}\ }\textbf {\bibinfo {volume} {327}},\ \bibinfo {pages} {1352}
  (\bibinfo {year} {2010})}\BibitemShut {NoStop}%
\bibitem [{\citenamefont {Rengstl}\ \emph {et~al.}(2015)\citenamefont
  {Rengstl}, \citenamefont {Schwartz}, \citenamefont {Herzog}, \citenamefont
  {Hargart}, \citenamefont {Paul}, \citenamefont {Portalupi}, \citenamefont
  {Jetter},\ and\ \citenamefont {Michler}}]{23}%
  \BibitemOpen
  \bibfield  {author} {\bibinfo {author} {\bibfnamefont {U.}~\bibnamefont
  {Rengstl}}, \bibinfo {author} {\bibfnamefont {M.}~\bibnamefont {Schwartz}},
  \bibinfo {author} {\bibfnamefont {T.}~\bibnamefont {Herzog}}, \bibinfo
  {author} {\bibfnamefont {F.}~\bibnamefont {Hargart}}, \bibinfo {author}
  {\bibfnamefont {M.}~\bibnamefont {Paul}}, \bibinfo {author} {\bibfnamefont
  {S.~L.}\ \bibnamefont {Portalupi}}, \bibinfo {author} {\bibfnamefont
  {M.}~\bibnamefont {Jetter}}, \ and\ \bibinfo {author} {\bibfnamefont
  {P.}~\bibnamefont {Michler}},\ }\href {https://doi.org/10.1063/1.4926729}
  {\bibfield  {journal} {\bibinfo  {journal} {Applied Physics Letters}\
  }\textbf {\bibinfo {volume} {107}},\ \bibinfo {pages} {021101} (\bibinfo
  {year} {2015})}\BibitemShut {NoStop}%
\bibitem [{\citenamefont {Makhonin}\ \emph {et~al.}(2014)\citenamefont
  {Makhonin}, \citenamefont {Dixon}, \citenamefont {Coles}, \citenamefont
  {Royall}, \citenamefont {Luxmoore}, \citenamefont {Clarke}, \citenamefont
  {Hugues}, \citenamefont {Skolnick},\ and\ \citenamefont {Fox}}]{30}%
  \BibitemOpen
  \bibfield  {author} {\bibinfo {author} {\bibfnamefont {M.~N.}\ \bibnamefont
  {Makhonin}}, \bibinfo {author} {\bibfnamefont {J.~E.}\ \bibnamefont {Dixon}},
  \bibinfo {author} {\bibfnamefont {R.~J.}\ \bibnamefont {Coles}}, \bibinfo
  {author} {\bibfnamefont {B.}~\bibnamefont {Royall}}, \bibinfo {author}
  {\bibfnamefont {I.~J.}\ \bibnamefont {Luxmoore}}, \bibinfo {author}
  {\bibfnamefont {E.}~\bibnamefont {Clarke}}, \bibinfo {author} {\bibfnamefont
  {M.}~\bibnamefont {Hugues}}, \bibinfo {author} {\bibfnamefont {M.~S.}\
  \bibnamefont {Skolnick}}, \ and\ \bibinfo {author} {\bibfnamefont {A.~M.}\
  \bibnamefont {Fox}},\ }\href {\doibase 10.1021/nl5032937} {\bibfield
  {journal} {\bibinfo  {journal} {Nano Letters}\ }\textbf {\bibinfo {volume}
  {14}},\ \bibinfo {pages} {6997} (\bibinfo {year} {2014})}\BibitemShut
  {NoStop}%
\bibitem [{\citenamefont {Schwartz}\ \emph {et~al.}(2016)\citenamefont
  {Schwartz}, \citenamefont {Rengstl}, \citenamefont {Herzog}, \citenamefont
  {Paul}, \citenamefont {Kettler}, \citenamefont {Portalupi}, \citenamefont
  {Jetter},\ and\ \citenamefont {Michler}}]{31}%
  \BibitemOpen
  \bibfield  {author} {\bibinfo {author} {\bibfnamefont {M.}~\bibnamefont
  {Schwartz}}, \bibinfo {author} {\bibfnamefont {U.}~\bibnamefont {Rengstl}},
  \bibinfo {author} {\bibfnamefont {T.}~\bibnamefont {Herzog}}, \bibinfo
  {author} {\bibfnamefont {M.}~\bibnamefont {Paul}}, \bibinfo {author}
  {\bibfnamefont {J.}~\bibnamefont {Kettler}}, \bibinfo {author} {\bibfnamefont
  {S.~L.}\ \bibnamefont {Portalupi}}, \bibinfo {author} {\bibfnamefont
  {M.}~\bibnamefont {Jetter}}, \ and\ \bibinfo {author} {\bibfnamefont
  {P.}~\bibnamefont {Michler}},\ }\href {\doibase 10.1364/OE.24.003089}
  {\bibfield  {journal} {\bibinfo  {journal} {Opt. Express}\ }\textbf {\bibinfo
  {volume} {24}},\ \bibinfo {pages} {3089} (\bibinfo {year}
  {2016})}\BibitemShut {NoStop}%
\bibitem [{\citenamefont {Schnauber}\ \emph {et~al.}(2018)\citenamefont
  {Schnauber}, \citenamefont {Schall}, \citenamefont {Bounouar}, \citenamefont
  {Höhne}, \citenamefont {Park}, \citenamefont {Ryu}, \citenamefont {Heindel},
  \citenamefont {Burger}, \citenamefont {Song}, \citenamefont {Rodt},\ and\
  \citenamefont {Reitzenstein}}]{Schauber.Schall.ea.2018}%
  \BibitemOpen
  \bibfield  {author} {\bibinfo {author} {\bibfnamefont {P.}~\bibnamefont
  {Schnauber}}, \bibinfo {author} {\bibfnamefont {J.}~\bibnamefont {Schall}},
  \bibinfo {author} {\bibfnamefont {S.}~\bibnamefont {Bounouar}}, \bibinfo
  {author} {\bibfnamefont {T.}~\bibnamefont {Höhne}}, \bibinfo {author}
  {\bibfnamefont {S.-I.}\ \bibnamefont {Park}}, \bibinfo {author}
  {\bibfnamefont {G.-H.}\ \bibnamefont {Ryu}}, \bibinfo {author} {\bibfnamefont
  {T.}~\bibnamefont {Heindel}}, \bibinfo {author} {\bibfnamefont
  {S.}~\bibnamefont {Burger}}, \bibinfo {author} {\bibfnamefont {J.-D.}\
  \bibnamefont {Song}}, \bibinfo {author} {\bibfnamefont {S.}~\bibnamefont
  {Rodt}}, \ and\ \bibinfo {author} {\bibfnamefont {S.}~\bibnamefont
  {Reitzenstein}},\ }\href {\doibase 10.1021/acs.nanolett.7b05218} {\bibfield
  {journal} {\bibinfo  {journal} {Nano Letters}\ }\textbf {\bibinfo {volume}
  {18}},\ \bibinfo {pages} {2336} (\bibinfo {year} {2018})}\BibitemShut
  {NoStop}%
\bibitem [{\citenamefont {Sprengers}\ \emph {et~al.}(2011)\citenamefont
  {Sprengers}, \citenamefont {Gaggero}, \citenamefont {Sahin}, \citenamefont
  {Jahanmirinejad}, \citenamefont {Frucci}, \citenamefont {Mattioli},
  \citenamefont {Leoni}, \citenamefont {Beetz}, \citenamefont {Lermer},
  \citenamefont {Kamp}, \citenamefont {H\"ofling}, \citenamefont {Sanjines},\
  and\ \citenamefont {Fiore}}]{WGSNSPD}%
  \BibitemOpen
  \bibfield  {author} {\bibinfo {author} {\bibfnamefont {J.~P.}\ \bibnamefont
  {Sprengers}}, \bibinfo {author} {\bibfnamefont {A.}~\bibnamefont {Gaggero}},
  \bibinfo {author} {\bibfnamefont {D.}~\bibnamefont {Sahin}}, \bibinfo
  {author} {\bibfnamefont {S.}~\bibnamefont {Jahanmirinejad}}, \bibinfo
  {author} {\bibfnamefont {G.}~\bibnamefont {Frucci}}, \bibinfo {author}
  {\bibfnamefont {F.}~\bibnamefont {Mattioli}}, \bibinfo {author}
  {\bibfnamefont {R.}~\bibnamefont {Leoni}}, \bibinfo {author} {\bibfnamefont
  {J.}~\bibnamefont {Beetz}}, \bibinfo {author} {\bibfnamefont
  {M.}~\bibnamefont {Lermer}}, \bibinfo {author} {\bibfnamefont
  {M.}~\bibnamefont {Kamp}}, \bibinfo {author} {\bibfnamefont {S.}~\bibnamefont
  {H\"ofling}}, \bibinfo {author} {\bibfnamefont {R.}~\bibnamefont {Sanjines}},
  \ and\ \bibinfo {author} {\bibfnamefont {A.}~\bibnamefont {Fiore}},\ }\href
  {https://doi.org/10.1063/1.3657518} {\bibfield  {journal} {\bibinfo
  {journal} {Applied Physics Letters}\ }\textbf {\bibinfo {volume} {99}},\
  \bibinfo {pages} {181110} (\bibinfo {year} {2011})}\BibitemShut {NoStop}%
\bibitem [{\citenamefont {Digeronimo}\ \emph {et~al.}(2016)\citenamefont
  {Digeronimo}, \citenamefont {Petruzzella}, \citenamefont {Birindelli},
  \citenamefont {Gaudio}, \citenamefont {Fattah~Poor}, \citenamefont {van
  Otten},\ and\ \citenamefont {Fiore}}]{12}%
  \BibitemOpen
  \bibfield  {author} {\bibinfo {author} {\bibfnamefont {G.~E.}\ \bibnamefont
  {Digeronimo}}, \bibinfo {author} {\bibfnamefont {M.}~\bibnamefont
  {Petruzzella}}, \bibinfo {author} {\bibfnamefont {S.}~\bibnamefont
  {Birindelli}}, \bibinfo {author} {\bibfnamefont {R.}~\bibnamefont {Gaudio}},
  \bibinfo {author} {\bibfnamefont {S.}~\bibnamefont {Fattah~Poor}}, \bibinfo
  {author} {\bibfnamefont {F.~W.}\ \bibnamefont {van Otten}}, \ and\ \bibinfo
  {author} {\bibfnamefont {A.}~\bibnamefont {Fiore}},\ }in\ \href
  {http://www.mdpi.com/2304-6732/3/4/55} {\emph {\bibinfo {booktitle}
  {Photonics}}},\ Vol.~\bibinfo {volume} {3}\ (\bibinfo {organization}
  {Multidisciplinary Digital Publishing Institute},\ \bibinfo {year} {2016})\
  p.~\bibinfo {pages} {55}\BibitemShut {NoStop}%
\bibitem [{\citenamefont {Marsili}\ \emph {et~al.}(2013)\citenamefont
  {Marsili}, \citenamefont {Verma}, \citenamefont {Stern}, \citenamefont
  {Harrington}, \citenamefont {Lita}, \citenamefont {Gerrits}, \citenamefont
  {Vayshenker}, \citenamefont {Baek}, \citenamefont {Shaw}, \citenamefont
  {Mirin},\ and\ \citenamefont {Nam}}]{marsili2013detecting}%
  \BibitemOpen
  \bibfield  {author} {\bibinfo {author} {\bibfnamefont {F.}~\bibnamefont
  {Marsili}}, \bibinfo {author} {\bibfnamefont {V.~B.}\ \bibnamefont {Verma}},
  \bibinfo {author} {\bibfnamefont {J.~A.}\ \bibnamefont {Stern}}, \bibinfo
  {author} {\bibfnamefont {S.}~\bibnamefont {Harrington}}, \bibinfo {author}
  {\bibfnamefont {A.~E.}\ \bibnamefont {Lita}}, \bibinfo {author}
  {\bibfnamefont {T.}~\bibnamefont {Gerrits}}, \bibinfo {author} {\bibfnamefont
  {I.}~\bibnamefont {Vayshenker}}, \bibinfo {author} {\bibfnamefont
  {B.}~\bibnamefont {Baek}}, \bibinfo {author} {\bibfnamefont {M.~D.}\
  \bibnamefont {Shaw}}, \bibinfo {author} {\bibfnamefont {R.~P.}\ \bibnamefont
  {Mirin}}, \ and\ \bibinfo {author} {\bibfnamefont {S.~W.}\ \bibnamefont
  {Nam}},\ }\href {http://dx.doi.org/10.1038/nphoton.2013.13} {\bibfield
  {journal} {\bibinfo  {journal} {Nature Photonics}\ }\textbf {\bibinfo
  {volume} {7}},\ \bibinfo {pages} {210} (\bibinfo {year} {2013})}\BibitemShut
  {NoStop}%
\bibitem [{\citenamefont {Sidorova}\ \emph {et~al.}(2017)\citenamefont
  {Sidorova}, \citenamefont {Semenov}, \citenamefont {H\"ubers}, \citenamefont
  {Charaev}, \citenamefont {Kuzmin}, \citenamefont {Doerner},\ and\
  \citenamefont {Siegel}}]{Jitter1}%
  \BibitemOpen
  \bibfield  {author} {\bibinfo {author} {\bibfnamefont {M.}~\bibnamefont
  {Sidorova}}, \bibinfo {author} {\bibfnamefont {A.}~\bibnamefont {Semenov}},
  \bibinfo {author} {\bibfnamefont {H.-W.}\ \bibnamefont {H\"ubers}}, \bibinfo
  {author} {\bibfnamefont {I.}~\bibnamefont {Charaev}}, \bibinfo {author}
  {\bibfnamefont {A.}~\bibnamefont {Kuzmin}}, \bibinfo {author} {\bibfnamefont
  {S.}~\bibnamefont {Doerner}}, \ and\ \bibinfo {author} {\bibfnamefont
  {M.}~\bibnamefont {Siegel}},\ }\href {\doibase 10.1103/PhysRevB.96.184504}
  {\bibfield  {journal} {\bibinfo  {journal} {Phys. Rev. B}\ }\textbf {\bibinfo
  {volume} {96}},\ \bibinfo {pages} {184504} (\bibinfo {year}
  {2017})}\BibitemShut {NoStop}%
\bibitem [{\citenamefont {Korzh}\ \emph {et~al.}(2018)\citenamefont {Korzh},
  \citenamefont {Zhao}, \citenamefont {Frasca}, \citenamefont {Allmaras},
  \citenamefont {Autry}, \citenamefont {Bersin}, \citenamefont {Colangelo},
  \citenamefont {Crouch}, \citenamefont {Dane}, \citenamefont {Gerrits},
  \citenamefont {Marsili}, \citenamefont {Moody}, \citenamefont {Ramirez},
  \citenamefont {Rezac}, \citenamefont {Stevens}, \citenamefont {Wollman},
  \citenamefont {Zhu}, \citenamefont {Hale}, \citenamefont {Silverman},
  \citenamefont {Minin}, \citenamefont {Nam}, \citenamefont {Shaw},\ and\
  \citenamefont {Berggren}}]{Jitter2}%
  \BibitemOpen
  \bibfield  {author} {\bibinfo {author} {\bibfnamefont {B.~A.}\ \bibnamefont
  {Korzh}}, \bibinfo {author} {\bibfnamefont {Q.~Y.}\ \bibnamefont {Zhao}},
  \bibinfo {author} {\bibfnamefont {S.}~\bibnamefont {Frasca}}, \bibinfo
  {author} {\bibfnamefont {J.~P.}\ \bibnamefont {Allmaras}}, \bibinfo {author}
  {\bibfnamefont {T.~M.}\ \bibnamefont {Autry}}, \bibinfo {author}
  {\bibfnamefont {E.~A.}\ \bibnamefont {Bersin}}, \bibinfo {author}
  {\bibfnamefont {M.}~\bibnamefont {Colangelo}}, \bibinfo {author}
  {\bibfnamefont {G.~M.}\ \bibnamefont {Crouch}}, \bibinfo {author}
  {\bibfnamefont {A.~E.}\ \bibnamefont {Dane}}, \bibinfo {author}
  {\bibfnamefont {T.}~\bibnamefont {Gerrits}}, \bibinfo {author} {\bibfnamefont
  {F.}~\bibnamefont {Marsili}}, \bibinfo {author} {\bibfnamefont
  {G.}~\bibnamefont {Moody}}, \bibinfo {author} {\bibfnamefont
  {E.}~\bibnamefont {Ramirez}}, \bibinfo {author} {\bibfnamefont {J.~D.}\
  \bibnamefont {Rezac}}, \bibinfo {author} {\bibfnamefont {M.~J.}\ \bibnamefont
  {Stevens}}, \bibinfo {author} {\bibfnamefont {E.~E.}\ \bibnamefont
  {Wollman}}, \bibinfo {author} {\bibfnamefont {D.}~\bibnamefont {Zhu}},
  \bibinfo {author} {\bibfnamefont {P.~D.}\ \bibnamefont {Hale}}, \bibinfo
  {author} {\bibfnamefont {K.~L.}\ \bibnamefont {Silverman}}, \bibinfo {author}
  {\bibfnamefont {R.~P.}\ \bibnamefont {Minin}}, \bibinfo {author}
  {\bibfnamefont {S.~W.}\ \bibnamefont {Nam}}, \bibinfo {author} {\bibfnamefont
  {M.~D.}\ \bibnamefont {Shaw}}, \ and\ \bibinfo {author} {\bibfnamefont
  {K.~K.}\ \bibnamefont {Berggren}},\ }\href {https://arxiv.org/abs/1804.06839}
  {\bibfield  {journal} {\bibinfo  {journal} {arXiv preprint arXiv:1804.06839}\
  } (\bibinfo {year} {2018})}\BibitemShut {NoStop}%
\bibitem [{\citenamefont {Bentham}\ \emph {et~al.}(2016)\citenamefont
  {Bentham}, \citenamefont {Hallett}, \citenamefont {Prtljaga}, \citenamefont
  {Royall}, \citenamefont {Vaitiekus}, \citenamefont {Coles}, \citenamefont
  {Clarke}, \citenamefont {Fox}, \citenamefont {Skolnick}, \citenamefont
  {Itskevich},\ and\ \citenamefont {Wilson}}]{Benthal.Hallett.ea.2016}%
  \BibitemOpen
  \bibfield  {author} {\bibinfo {author} {\bibfnamefont {C.}~\bibnamefont
  {Bentham}}, \bibinfo {author} {\bibfnamefont {D.}~\bibnamefont {Hallett}},
  \bibinfo {author} {\bibfnamefont {N.}~\bibnamefont {Prtljaga}}, \bibinfo
  {author} {\bibfnamefont {B.}~\bibnamefont {Royall}}, \bibinfo {author}
  {\bibfnamefont {D.}~\bibnamefont {Vaitiekus}}, \bibinfo {author}
  {\bibfnamefont {R.~J.}\ \bibnamefont {Coles}}, \bibinfo {author}
  {\bibfnamefont {E.}~\bibnamefont {Clarke}}, \bibinfo {author} {\bibfnamefont
  {A.~M.}\ \bibnamefont {Fox}}, \bibinfo {author} {\bibfnamefont {M.~S.}\
  \bibnamefont {Skolnick}}, \bibinfo {author} {\bibfnamefont {I.~E.}\
  \bibnamefont {Itskevich}}, \ and\ \bibinfo {author} {\bibfnamefont {L.~R.}\
  \bibnamefont {Wilson}},\ }\href {\doibase 10.1063/1.4965295} {\bibfield
  {journal} {\bibinfo  {journal} {Applied Physics Letters}\ }\textbf {\bibinfo
  {volume} {109}},\ \bibinfo {pages} {161101} (\bibinfo {year}
  {2016})}\BibitemShut {NoStop}%
\bibitem [{\citenamefont {Zadeh}\ \emph {et~al.}(2016)\citenamefont {Zadeh},
  \citenamefont {Elshaari}, \citenamefont {J\"ons}, \citenamefont {Fognini},
  \citenamefont {Dalacu}, \citenamefont {Poole}, \citenamefont {Reimer},\ and\
  \citenamefont {Zwiller}}]{detQDonSi}%
  \BibitemOpen
  \bibfield  {author} {\bibinfo {author} {\bibfnamefont {I.~E.}\ \bibnamefont
  {Zadeh}}, \bibinfo {author} {\bibfnamefont {A.~W.}\ \bibnamefont {Elshaari}},
  \bibinfo {author} {\bibfnamefont {K.~D.}\ \bibnamefont {J\"ons}}, \bibinfo
  {author} {\bibfnamefont {A.}~\bibnamefont {Fognini}}, \bibinfo {author}
  {\bibfnamefont {D.}~\bibnamefont {Dalacu}}, \bibinfo {author} {\bibfnamefont
  {P.~J.}\ \bibnamefont {Poole}}, \bibinfo {author} {\bibfnamefont {M.~E.}\
  \bibnamefont {Reimer}}, \ and\ \bibinfo {author} {\bibfnamefont
  {V.}~\bibnamefont {Zwiller}},\ }\href {\doibase 10.1021/acs.nanolett.5b04709}
  {\bibfield  {journal} {\bibinfo  {journal} {Nano Letters}\ }\textbf {\bibinfo
  {volume} {16}},\ \bibinfo {pages} {2289} (\bibinfo {year}
  {2016})}\BibitemShut {NoStop}%
\bibitem [{\citenamefont {Davanco}\ \emph {et~al.}(2017)\citenamefont
  {Davanco}, \citenamefont {Liu}, \citenamefont {Sapienza}, \citenamefont
  {Zhang}, \citenamefont {Cardoso}, \citenamefont {Verma}, \citenamefont
  {Mirin}, \citenamefont {Nam}, \citenamefont {Liu},\ and\ \citenamefont
  {Srinivasan}}]{hetPICQD}%
  \BibitemOpen
  \bibfield  {author} {\bibinfo {author} {\bibfnamefont {M.}~\bibnamefont
  {Davanco}}, \bibinfo {author} {\bibfnamefont {J.}~\bibnamefont {Liu}},
  \bibinfo {author} {\bibfnamefont {L.}~\bibnamefont {Sapienza}}, \bibinfo
  {author} {\bibfnamefont {C.-Z.}\ \bibnamefont {Zhang}}, \bibinfo {author}
  {\bibfnamefont {J.~V.~M.}\ \bibnamefont {Cardoso}}, \bibinfo {author}
  {\bibfnamefont {V.}~\bibnamefont {Verma}}, \bibinfo {author} {\bibfnamefont
  {R.}~\bibnamefont {Mirin}}, \bibinfo {author} {\bibfnamefont {S.~W.}\
  \bibnamefont {Nam}}, \bibinfo {author} {\bibfnamefont {L.}~\bibnamefont
  {Liu}}, \ and\ \bibinfo {author} {\bibfnamefont {K.}~\bibnamefont
  {Srinivasan}},\ }\href {https://doi.org/10.1038/s41467-017-00987-6}
  {\bibfield  {journal} {\bibinfo  {journal} {Nature communications}\ }\textbf
  {\bibinfo {volume} {8}},\ \bibinfo {pages} {889} (\bibinfo {year}
  {2017})}\BibitemShut {NoStop}%
\bibitem [{\citenamefont {Murray}\ \emph {et~al.}(2015)\citenamefont {Murray},
  \citenamefont {Ellis}, \citenamefont {Meany}, \citenamefont {Floether},
  \citenamefont {Lee}, \citenamefont {Griffiths}, \citenamefont {Jones},
  \citenamefont {Farrer}, \citenamefont {Ritchie}, \citenamefont {Bennett},\
  and\ \citenamefont {Shields}}]{Murray.Ellis.ea.2015}%
  \BibitemOpen
  \bibfield  {author} {\bibinfo {author} {\bibfnamefont {E.}~\bibnamefont
  {Murray}}, \bibinfo {author} {\bibfnamefont {D.~J.~P.}\ \bibnamefont
  {Ellis}}, \bibinfo {author} {\bibfnamefont {T.}~\bibnamefont {Meany}},
  \bibinfo {author} {\bibfnamefont {F.~F.}\ \bibnamefont {Floether}}, \bibinfo
  {author} {\bibfnamefont {J.~P.}\ \bibnamefont {Lee}}, \bibinfo {author}
  {\bibfnamefont {J.~P.}\ \bibnamefont {Griffiths}}, \bibinfo {author}
  {\bibfnamefont {G.~A.~C.}\ \bibnamefont {Jones}}, \bibinfo {author}
  {\bibfnamefont {I.}~\bibnamefont {Farrer}}, \bibinfo {author} {\bibfnamefont
  {D.~A.}\ \bibnamefont {Ritchie}}, \bibinfo {author} {\bibfnamefont {A.~J.}\
  \bibnamefont {Bennett}}, \ and\ \bibinfo {author} {\bibfnamefont {A.~J.}\
  \bibnamefont {Shields}},\ }\href {\doibase 10.1063/1.4935029} {\bibfield
  {journal} {\bibinfo  {journal} {Applied Physics Letters}\ }\textbf {\bibinfo
  {volume} {107}},\ \bibinfo {pages} {171108} (\bibinfo {year}
  {2015})}\BibitemShut {NoStop}%
\bibitem [{\citenamefont {Ellis}\ \emph {et~al.}(2018)\citenamefont {Ellis},
  \citenamefont {Bennett}, \citenamefont {Dangel}, \citenamefont {Lee},
  \citenamefont {Griffiths}, \citenamefont {Mitchell}, \citenamefont {Paraiso},
  \citenamefont {Spencer}, \citenamefont {Ritchie},\ and\ \citenamefont
  {Shields}}]{Ellis.Bennett.ea.2018}%
  \BibitemOpen
  \bibfield  {author} {\bibinfo {author} {\bibfnamefont {D.~J.~P.}\
  \bibnamefont {Ellis}}, \bibinfo {author} {\bibfnamefont {A.~J.}\ \bibnamefont
  {Bennett}}, \bibinfo {author} {\bibfnamefont {C.}~\bibnamefont {Dangel}},
  \bibinfo {author} {\bibfnamefont {J.~P.}\ \bibnamefont {Lee}}, \bibinfo
  {author} {\bibfnamefont {J.~P.}\ \bibnamefont {Griffiths}}, \bibinfo {author}
  {\bibfnamefont {T.~A.}\ \bibnamefont {Mitchell}}, \bibinfo {author}
  {\bibfnamefont {T.-K.}\ \bibnamefont {Paraiso}}, \bibinfo {author}
  {\bibfnamefont {P.}~\bibnamefont {Spencer}}, \bibinfo {author} {\bibfnamefont
  {D.~A.}\ \bibnamefont {Ritchie}}, \ and\ \bibinfo {author} {\bibfnamefont
  {A.~J.}\ \bibnamefont {Shields}},\ }\href {\doibase 10.1063/1.5028339}
  {\bibfield  {journal} {\bibinfo  {journal} {Applied Physics Letters}\
  }\textbf {\bibinfo {volume} {112}},\ \bibinfo {pages} {211104} (\bibinfo
  {year} {2018})}\BibitemShut {NoStop}%
\bibitem [{\citenamefont {Richter}\ \emph {et~al.}(2010)\citenamefont
  {Richter}, \citenamefont {Hafenbrak}, \citenamefont {J{\"o}ns}, \citenamefont
  {Schulz}, \citenamefont {Eichfelder}, \citenamefont {Heldmaier},
  \citenamefont {Ro{\ss}bach}, \citenamefont {Jetter},\ and\ \citenamefont
  {Michler}}]{richter2010low}%
  \BibitemOpen
  \bibfield  {author} {\bibinfo {author} {\bibfnamefont {D.}~\bibnamefont
  {Richter}}, \bibinfo {author} {\bibfnamefont {R.}~\bibnamefont {Hafenbrak}},
  \bibinfo {author} {\bibfnamefont {K.~D.}\ \bibnamefont {J{\"o}ns}}, \bibinfo
  {author} {\bibfnamefont {W.-M.}\ \bibnamefont {Schulz}}, \bibinfo {author}
  {\bibfnamefont {M.}~\bibnamefont {Eichfelder}}, \bibinfo {author}
  {\bibfnamefont {M.}~\bibnamefont {Heldmaier}}, \bibinfo {author}
  {\bibfnamefont {R.}~\bibnamefont {Ro{\ss}bach}}, \bibinfo {author}
  {\bibfnamefont {M.}~\bibnamefont {Jetter}}, \ and\ \bibinfo {author}
  {\bibfnamefont {P.}~\bibnamefont {Michler}},\ }\href
  {http://stacks.iop.org/0957-4484/21/i=12/a=125606} {\bibfield  {journal}
  {\bibinfo  {journal} {Nanotechnology}\ }\textbf {\bibinfo {volume} {21}},\
  \bibinfo {pages} {125606} (\bibinfo {year} {2010})}\BibitemShut {NoStop}%
\bibitem [{\citenamefont {Fischer}\ \emph {et~al.}(2016)\citenamefont
  {Fischer}, \citenamefont {M{\"u}ller}, \citenamefont {Lagoudakis},\ and\
  \citenamefont {Vu{\v{c}}kovi{\'c}}}]{fischer2016dynamical}%
  \BibitemOpen
  \bibfield  {author} {\bibinfo {author} {\bibfnamefont {K.~A.}\ \bibnamefont
  {Fischer}}, \bibinfo {author} {\bibfnamefont {K.}~\bibnamefont {M{\"u}ller}},
  \bibinfo {author} {\bibfnamefont {K.~G.}\ \bibnamefont {Lagoudakis}}, \ and\
  \bibinfo {author} {\bibfnamefont {J.}~\bibnamefont {Vu{\v{c}}kovi{\'c}}},\
  }\href {http://stacks.iop.org/1367-2630/18/i=11/a=113053} {\bibfield
  {journal} {\bibinfo  {journal} {New Journal of Physics}\ }\textbf {\bibinfo
  {volume} {18}},\ \bibinfo {pages} {113053} (\bibinfo {year}
  {2016})}\BibitemShut {NoStop}%
\bibitem [{\citenamefont {Flissikowski}\ \emph {et~al.}(2001)\citenamefont
  {Flissikowski}, \citenamefont {Hundt}, \citenamefont {Lowisch}, \citenamefont
  {Rabe},\ and\ \citenamefont {Henneberger}}]{PhotonBeats}%
  \BibitemOpen
  \bibfield  {author} {\bibinfo {author} {\bibfnamefont {T.}~\bibnamefont
  {Flissikowski}}, \bibinfo {author} {\bibfnamefont {A.}~\bibnamefont {Hundt}},
  \bibinfo {author} {\bibfnamefont {M.}~\bibnamefont {Lowisch}}, \bibinfo
  {author} {\bibfnamefont {M.}~\bibnamefont {Rabe}}, \ and\ \bibinfo {author}
  {\bibfnamefont {F.}~\bibnamefont {Henneberger}},\ }\href {\doibase
  10.1103/PhysRevLett.86.3172} {\bibfield  {journal} {\bibinfo  {journal}
  {Phys. Rev. Lett.}\ }\textbf {\bibinfo {volume} {86}},\ \bibinfo {pages}
  {3172} (\bibinfo {year} {2001})}\BibitemShut {NoStop}%
\bibitem [{\citenamefont {Kitaygorsky}\ \emph {et~al.}(2007)\citenamefont
  {Kitaygorsky}, \citenamefont {Komissarov}, \citenamefont {Jukna},
  \citenamefont {Pan}, \citenamefont {Minaeva}, \citenamefont {Kaurova},
  \citenamefont {Divochiy}, \citenamefont {Korneev}, \citenamefont {Tarkhov},
  \citenamefont {Voronov}, \citenamefont {Milostnaya}, \citenamefont
  {Gol'tsman},\ and\ \citenamefont {Sobolewski}}]{kitaygorsky2007dark}%
  \BibitemOpen
  \bibfield  {author} {\bibinfo {author} {\bibfnamefont {J.}~\bibnamefont
  {Kitaygorsky}}, \bibinfo {author} {\bibfnamefont {I.}~\bibnamefont
  {Komissarov}}, \bibinfo {author} {\bibfnamefont {A.}~\bibnamefont {Jukna}},
  \bibinfo {author} {\bibfnamefont {D.}~\bibnamefont {Pan}}, \bibinfo {author}
  {\bibfnamefont {O.}~\bibnamefont {Minaeva}}, \bibinfo {author} {\bibfnamefont
  {N.}~\bibnamefont {Kaurova}}, \bibinfo {author} {\bibfnamefont
  {A.}~\bibnamefont {Divochiy}}, \bibinfo {author} {\bibfnamefont
  {A.}~\bibnamefont {Korneev}}, \bibinfo {author} {\bibfnamefont
  {M.}~\bibnamefont {Tarkhov}}, \bibinfo {author} {\bibfnamefont
  {B.}~\bibnamefont {Voronov}}, \bibinfo {author} {\bibfnamefont
  {I.}~\bibnamefont {Milostnaya}}, \bibinfo {author} {\bibfnamefont
  {G.}~\bibnamefont {Gol'tsman}}, \ and\ \bibinfo {author} {\bibfnamefont
  {R.~R.}\ \bibnamefont {Sobolewski}},\ }\href {\doibase
  10.1109/TASC.2007.898109} {\bibfield  {journal} {\bibinfo  {journal} {IEEE
  Transactions on Applied Superconductivity}\ }\textbf {\bibinfo {volume}
  {17}},\ \bibinfo {pages} {275} (\bibinfo {year} {2007})}\BibitemShut
  {NoStop}%
\bibitem [{\citenamefont {Patel}\ \emph {et~al.}(2010)\citenamefont {Patel},
  \citenamefont {Bennett}, \citenamefont {Farrer}, \citenamefont {Nicoll},
  \citenamefont {Ritchie},\ and\ \citenamefont
  {Shields}}]{Patel.Bennett.ea.2010}%
  \BibitemOpen
  \bibfield  {author} {\bibinfo {author} {\bibfnamefont {R.~B.}\ \bibnamefont
  {Patel}}, \bibinfo {author} {\bibfnamefont {A.~J.}\ \bibnamefont {Bennett}},
  \bibinfo {author} {\bibfnamefont {I.}~\bibnamefont {Farrer}}, \bibinfo
  {author} {\bibfnamefont {C.~A.}\ \bibnamefont {Nicoll}}, \bibinfo {author}
  {\bibfnamefont {D.~A.}\ \bibnamefont {Ritchie}}, \ and\ \bibinfo {author}
  {\bibfnamefont {A.~J.}\ \bibnamefont {Shields}},\ }\href
  {http://dx.doi.org/10.1038/nphoton.2010.161} {\bibfield  {journal} {\bibinfo
  {journal} {Nature Photonics}\ }\textbf {\bibinfo {volume} {4}},\ \bibinfo
  {pages} {632} (\bibinfo {year} {2010})}\BibitemShut {NoStop}%
\bibitem [{\citenamefont {Schmidt}\ \emph {et~al.}(2017)\citenamefont
  {Schmidt}, \citenamefont {Ilin},\ and\ \citenamefont
  {Siegel}}]{schmidt2017aln}%
  \BibitemOpen
  \bibfield  {author} {\bibinfo {author} {\bibfnamefont {E.}~\bibnamefont
  {Schmidt}}, \bibinfo {author} {\bibfnamefont {K.}~\bibnamefont {Ilin}}, \
  and\ \bibinfo {author} {\bibfnamefont {M.}~\bibnamefont {Siegel}},\ }\href
  {\doibase 10.1109/TASC.2016.2637358} {\bibfield  {journal} {\bibinfo
  {journal} {IEEE Transactions on Applied Superconductivity}\ }\textbf
  {\bibinfo {volume} {27}},\ \bibinfo {pages} {1} (\bibinfo {year}
  {2017})}\BibitemShut {NoStop}%
\bibitem [{\citenamefont {Clem}\ and\ \citenamefont
  {Berggren}(2011)}]{clem2011geometry}%
  \BibitemOpen
  \bibfield  {author} {\bibinfo {author} {\bibfnamefont {J.~R.}\ \bibnamefont
  {Clem}}\ and\ \bibinfo {author} {\bibfnamefont {K.~K.}\ \bibnamefont
  {Berggren}},\ }\href {\doibase 10.1103/PhysRevB.84.174510} {\bibfield
  {journal} {\bibinfo  {journal} {Physical Review B}\ }\textbf {\bibinfo
  {volume} {84}},\ \bibinfo {pages} {174510} (\bibinfo {year}
  {2011})}\BibitemShut {NoStop}%
\bibitem [{\citenamefont {Henrich}\ \emph {et~al.}(2013)\citenamefont
  {Henrich}, \citenamefont {Rehm}, \citenamefont {D{\"o}rner}, \citenamefont
  {Hofherr}, \citenamefont {Il'in}, \citenamefont {Semenov},\ and\
  \citenamefont {Siegel}}]{henrich2013detection}%
  \BibitemOpen
  \bibfield  {author} {\bibinfo {author} {\bibfnamefont {D.}~\bibnamefont
  {Henrich}}, \bibinfo {author} {\bibfnamefont {L.}~\bibnamefont {Rehm}},
  \bibinfo {author} {\bibfnamefont {S.}~\bibnamefont {D{\"o}rner}}, \bibinfo
  {author} {\bibfnamefont {M.}~\bibnamefont {Hofherr}}, \bibinfo {author}
  {\bibfnamefont {K.}~\bibnamefont {Il'in}}, \bibinfo {author} {\bibfnamefont
  {A.}~\bibnamefont {Semenov}}, \ and\ \bibinfo {author} {\bibfnamefont
  {M.}~\bibnamefont {Siegel}},\ }\href {\doibase 10.1109/TASC.2013.2237936}
  {\bibfield  {journal} {\bibinfo  {journal} {IEEE Transactions on Applied
  Superconductivity}\ }\textbf {\bibinfo {volume} {23}},\ \bibinfo {pages}
  {2200405} (\bibinfo {year} {2013})}\BibitemShut {NoStop}%
\bibitem [{\citenamefont {Nguyen}\ \emph {et~al.}(2012)\citenamefont {Nguyen},
  \citenamefont {Sallen}, \citenamefont {Voisin}, \citenamefont {Roussignol},
  \citenamefont {Diederichs},\ and\ \citenamefont
  {Cassabois}}]{Nguyen.Sallen.ea.2012}%
  \BibitemOpen
  \bibfield  {author} {\bibinfo {author} {\bibfnamefont {H.~S.}\ \bibnamefont
  {Nguyen}}, \bibinfo {author} {\bibfnamefont {G.}~\bibnamefont {Sallen}},
  \bibinfo {author} {\bibfnamefont {C.}~\bibnamefont {Voisin}}, \bibinfo
  {author} {\bibfnamefont {P.}~\bibnamefont {Roussignol}}, \bibinfo {author}
  {\bibfnamefont {C.}~\bibnamefont {Diederichs}}, \ and\ \bibinfo {author}
  {\bibfnamefont {G.}~\bibnamefont {Cassabois}},\ }\href {\doibase
  10.1103/PhysRevLett.108.057401} {\bibfield  {journal} {\bibinfo  {journal}
  {Phys. Rev. Lett.}\ }\textbf {\bibinfo {volume} {108}},\ \bibinfo {pages}
  {057401} (\bibinfo {year} {2012})}\BibitemShut {NoStop}%
\end{thebibliography}

\begin{thebibliography}{6}%
\makeatletter
\providecommand \@ifxundefined [1]{%
 \@ifx{#1\undefined}
}%
\providecommand \@ifnum [1]{%
 \ifnum #1\expandafter \@firstoftwo
 \else \expandafter \@secondoftwo
 \fi
}%
\providecommand \@ifx [1]{%
 \ifx #1\expandafter \@firstoftwo
 \else \expandafter \@secondoftwo
 \fi
}%
\providecommand \natexlab [1]{#1}%
\providecommand \enquote  [1]{``#1''}%
\providecommand \bibnamefont  [1]{#1}%
\providecommand \bibfnamefont [1]{#1}%
\providecommand \citenamefont [1]{#1}%
\providecommand \href@noop [0]{\@secondoftwo}%
\providecommand \href [0]{\begingroup \@sanitize@url \@href}%
\providecommand \@href[1]{\@@startlink{#1}\@@href}%
\providecommand \@@href[1]{\endgroup#1\@@endlink}%
\providecommand \@sanitize@url [0]{\catcode `\\12\catcode `\$12\catcode
  `\&12\catcode `\#12\catcode `\^12\catcode `\_12\catcode `\%12\relax}%
\providecommand \@@startlink[1]{}%
\providecommand \@@endlink[0]{}%
\providecommand \url  [0]{\begingroup\@sanitize@url \@url }%
\providecommand \@url [1]{\endgroup\@href {#1}{\urlprefix }}%
\providecommand \urlprefix  [0]{URL }%
\providecommand \Eprint [0]{\href }%
\providecommand \doibase [0]{http://dx.doi.org/}%
\providecommand \selectlanguage [0]{\@gobble}%
\providecommand \bibinfo  [0]{\@secondoftwo}%
\providecommand \bibfield  [0]{\@secondoftwo}%
\providecommand \translation [1]{[#1]}%
\providecommand \BibitemOpen [0]{}%
\providecommand \bibitemStop [0]{}%
\providecommand \bibitemNoStop [0]{.\EOS\space}%
\providecommand \EOS [0]{\spacefactor3000\relax}%
\providecommand \BibitemShut  [1]{\csname bibitem#1\endcsname}%
\let\auto@bib@innerbib\@empty
\bibitem [{\citenamefont {Zhang}\ \emph {et~al.}(2014)\citenamefont {Zhang},
  \citenamefont {You}, \citenamefont {Liu}, \citenamefont {Zhang},
  \citenamefont {Zhang}, \citenamefont {Liu}, \citenamefont {Wu}, \citenamefont
  {He}, \citenamefont {Lv}, \citenamefont {Wang},\ and\ \citenamefont
  {Xie}}]{zhang2014characterization}%
  \BibitemOpen
  \bibfield  {author} {\bibinfo {author} {\bibfnamefont {L.}~\bibnamefont
  {Zhang}}, \bibinfo {author} {\bibfnamefont {L.}~\bibnamefont {You}}, \bibinfo
  {author} {\bibfnamefont {D.}~\bibnamefont {Liu}}, \bibinfo {author}
  {\bibfnamefont {W.}~\bibnamefont {Zhang}}, \bibinfo {author} {\bibfnamefont
  {L.}~\bibnamefont {Zhang}}, \bibinfo {author} {\bibfnamefont
  {X.}~\bibnamefont {Liu}}, \bibinfo {author} {\bibfnamefont {J.}~\bibnamefont
  {Wu}}, \bibinfo {author} {\bibfnamefont {Y.}~\bibnamefont {He}}, \bibinfo
  {author} {\bibfnamefont {C.}~\bibnamefont {Lv}}, \bibinfo {author}
  {\bibfnamefont {Z.}~\bibnamefont {Wang}}, \ and\ \bibinfo {author}
  {\bibfnamefont {X.}~\bibnamefont {Xie}},\ }\href
  {https://doi.org/10.1063/1.4881981} {\bibfield  {journal} {\bibinfo
  {journal} {Aip Advances}\ }\textbf {\bibinfo {volume} {4}},\ \bibinfo {pages}
  {067114} (\bibinfo {year} {2014})}\BibitemShut {NoStop}%
\bibitem [{\citenamefont {Yamashita}\ \emph {et~al.}(2011)\citenamefont
  {Yamashita}, \citenamefont {Miki}, \citenamefont {Makise}, \citenamefont
  {Qiu}, \citenamefont {Terai}, \citenamefont {Fujiwara}, \citenamefont
  {Sasaki},\ and\ \citenamefont {Wang}}]{yamashita2011origin}%
  \BibitemOpen
  \bibfield  {author} {\bibinfo {author} {\bibfnamefont {T.}~\bibnamefont
  {Yamashita}}, \bibinfo {author} {\bibfnamefont {S.}~\bibnamefont {Miki}},
  \bibinfo {author} {\bibfnamefont {K.}~\bibnamefont {Makise}}, \bibinfo
  {author} {\bibfnamefont {W.}~\bibnamefont {Qiu}}, \bibinfo {author}
  {\bibfnamefont {H.}~\bibnamefont {Terai}}, \bibinfo {author} {\bibfnamefont
  {M.}~\bibnamefont {Fujiwara}}, \bibinfo {author} {\bibfnamefont
  {M.}~\bibnamefont {Sasaki}}, \ and\ \bibinfo {author} {\bibfnamefont
  {Z.}~\bibnamefont {Wang}},\ }\href {https://doi.org/10.1063/1.3652908}
  {\bibfield  {journal} {\bibinfo  {journal} {Applied Physics Letters}\
  }\textbf {\bibinfo {volume} {99}},\ \bibinfo {pages} {161105} (\bibinfo
  {year} {2011})}\BibitemShut {NoStop}%
\bibitem [{\citenamefont {Flissikowski}\ \emph {et~al.}(2001)\citenamefont
  {Flissikowski}, \citenamefont {Hundt}, \citenamefont {Lowisch}, \citenamefont
  {Rabe},\ and\ \citenamefont {Henneberger}}]{Flissikowski.Hundt.ea.2001}%
  \BibitemOpen
  \bibfield  {author} {\bibinfo {author} {\bibfnamefont {T.}~\bibnamefont
  {Flissikowski}}, \bibinfo {author} {\bibfnamefont {A.}~\bibnamefont {Hundt}},
  \bibinfo {author} {\bibfnamefont {M.}~\bibnamefont {Lowisch}}, \bibinfo
  {author} {\bibfnamefont {M.}~\bibnamefont {Rabe}}, \ and\ \bibinfo {author}
  {\bibfnamefont {F.}~\bibnamefont {Henneberger}},\ }\href {\doibase
  10.1103/PhysRevLett.86.3172} {\bibfield  {journal} {\bibinfo  {journal}
  {Phys. Rev. Lett.}\ }\textbf {\bibinfo {volume} {86}},\ \bibinfo {pages}
  {3172} (\bibinfo {year} {2001})}\BibitemShut {NoStop}%
\bibitem [{\citenamefont {Rengstl}\ \emph {et~al.}(2015)\citenamefont
  {Rengstl}, \citenamefont {Schwartz}, \citenamefont {Herzog}, \citenamefont
  {Hargart}, \citenamefont {Paul}, \citenamefont {Portalupi}, \citenamefont
  {Jetter},\ and\ \citenamefont {Michler}}]{Rengstl.Schwartz.ea.2015}%
  \BibitemOpen
  \bibfield  {author} {\bibinfo {author} {\bibfnamefont {U.}~\bibnamefont
  {Rengstl}}, \bibinfo {author} {\bibfnamefont {M.}~\bibnamefont {Schwartz}},
  \bibinfo {author} {\bibfnamefont {T.}~\bibnamefont {Herzog}}, \bibinfo
  {author} {\bibfnamefont {F.}~\bibnamefont {Hargart}}, \bibinfo {author}
  {\bibfnamefont {M.}~\bibnamefont {Paul}}, \bibinfo {author} {\bibfnamefont
  {S.}~\bibnamefont {Portalupi}}, \bibinfo {author} {\bibfnamefont
  {M.}~\bibnamefont {Jetter}}, \ and\ \bibinfo {author} {\bibfnamefont
  {P.}~\bibnamefont {Michler}},\ }\href {https://doi.org/10.1063/1.4926729}
  {\bibfield  {journal} {\bibinfo  {journal} {Applied Physics Letters}\
  }\textbf {\bibinfo {volume} {107}},\ \bibinfo {pages} {021101} (\bibinfo
  {year} {2015})}\BibitemShut {NoStop}%
\bibitem [{\citenamefont {Makhonin}\ \emph {et~al.}(2014)\citenamefont
  {Makhonin}, \citenamefont {Dixon}, \citenamefont {Coles}, \citenamefont
  {Royall}, \citenamefont {Luxmoore}, \citenamefont {Clarke}, \citenamefont
  {Hugues}, \citenamefont {Skolnick},\ and\ \citenamefont
  {Fox}}]{Makhonin.Dixon.ea.2014}%
  \BibitemOpen
  \bibfield  {author} {\bibinfo {author} {\bibfnamefont {M.~N.}\ \bibnamefont
  {Makhonin}}, \bibinfo {author} {\bibfnamefont {J.~E.}\ \bibnamefont {Dixon}},
  \bibinfo {author} {\bibfnamefont {R.~J.}\ \bibnamefont {Coles}}, \bibinfo
  {author} {\bibfnamefont {B.}~\bibnamefont {Royall}}, \bibinfo {author}
  {\bibfnamefont {I.~J.}\ \bibnamefont {Luxmoore}}, \bibinfo {author}
  {\bibfnamefont {E.}~\bibnamefont {Clarke}}, \bibinfo {author} {\bibfnamefont
  {M.}~\bibnamefont {Hugues}}, \bibinfo {author} {\bibfnamefont {M.~S.}\
  \bibnamefont {Skolnick}}, \ and\ \bibinfo {author} {\bibfnamefont {A.~M.}\
  \bibnamefont {Fox}},\ }\href {\doibase 10.1021/nl5032937} {\bibfield
  {journal} {\bibinfo  {journal} {Nano Letters}\ }\textbf {\bibinfo {volume}
  {14}},\ \bibinfo {pages} {6997} (\bibinfo {year} {2014})},\ \bibinfo {note}
  {pMID: 25381734}\BibitemShut {NoStop}%
\bibitem [{\citenamefont {Nguyen}\ \emph {et~al.}(2012)\citenamefont {Nguyen},
  \citenamefont {Sallen}, \citenamefont {Voisin}, \citenamefont {Roussignol},
  \citenamefont {Diederichs},\ and\ \citenamefont
  {Cassabois}}]{Nguyen.Sallen.ea.20122}%
  \BibitemOpen
  \bibfield  {author} {\bibinfo {author} {\bibfnamefont {H.~S.}\ \bibnamefont
  {Nguyen}}, \bibinfo {author} {\bibfnamefont {G.}~\bibnamefont {Sallen}},
  \bibinfo {author} {\bibfnamefont {C.}~\bibnamefont {Voisin}}, \bibinfo
  {author} {\bibfnamefont {P.}~\bibnamefont {Roussignol}}, \bibinfo {author}
  {\bibfnamefont {C.}~\bibnamefont {Diederichs}}, \ and\ \bibinfo {author}
  {\bibfnamefont {G.}~\bibnamefont {Cassabois}},\ }\href {\doibase
  10.1103/PhysRevLett.108.057401} {\bibfield  {journal} {\bibinfo  {journal}
  {Phys. Rev. Lett.}\ }\textbf {\bibinfo {volume} {108}},\ \bibinfo {pages}
  {057401} (\bibinfo {year} {2012})}\BibitemShut {NoStop}%
\end{thebibliography}

\FloatBarrier 

\end{document}